\documentclass[a4paper,english,eqsecnum]{article}
\usepackage[T1]{fontenc}
\usepackage[utf8]{luainputenc}
\usepackage{xcolor}
\usepackage{pdfcolmk}
\usepackage{float}
\usepackage{amsmath}
\usepackage{amssymb}
\usepackage{graphicx}
\PassOptionsToPackage{normalem}{ulem}
\usepackage{ulem}

\makeatletter

\pdfpageheight\paperheight
\pdfpagewidth\paperwidth

\newcommand{\lyxmathsym}[1]{\ifmmode\begingroup\def\b@ld{bold}
  \text{\ifx\math@version\b@ld\bfseries\fi#1}\endgroup\else#1\fi}

\providecolor{lyxadded}{rgb}{0,0,1}
\providecolor{lyxdeleted}{rgb}{1,0,0}

\usepackage{a4wide}
\usepackage{psfrag}
\makeatletter
\renewcommand{\theequation}{\hbox{\normalsize\arabic{section}.\arabic{equation}}}
\@addtoreset{equation}{section}
\renewcommand{\thefigure}{\hbox{\normalsize\arabic{section}.\arabic{figure}}}
\@addtoreset{figure}{section}
\renewcommand{\thetable}{\hbox{\normalsize\arabic{section}.\arabic{table}}}
\@addtoreset{table}{section}
\makeatother

\usepackage{ifpdf}\ifpdf\usepackage{epstopdf}\usepackage[pdftex,ps2pdf,dvips,colorlinks,urlcolor=blue,citecolor=blue,linkcolor=blue]{hyperref}\else
\usepackage[hypertex,ps2pdf,dvips,colorlinks,urlcolor=blue,citecolor=blue,linkcolor=blue]{hyperref}\fi\pdfadjustspacing=1

\makeatother

\usepackage{babel}
\begin{document}

\title{Finite temperature one-point functions in non-diagonal integrable
field theories: the sine-Gordon model}

\author{F. Buccheri$^{1}$ and G. Takács$^{2,3}$\\
~\\
$^{1}$International Institute of Physics, Universidade Federal do
Rio Grande do Norte \\
Av. Odilon Gomes de Lima, 1722 - Natal-RN, Brazil\\
~\\
$^{2}$MTA-BME ``Momentum'' Statistical Field Theory Research Group\\
 1111 Budapest, Budafoki út 8, Hungary\\
~\\
$^{3}$Department of Theoretical Physics, \\
 Budapest University of Technology and Economics\\
 1111 Budapest, Budafoki út 8, Hungary\\
}

\date{8th December 2013}
\maketitle
\begin{abstract}
We study the finite-temperature expectation values of exponential
fields in the sine-Gordon model. Using finite-volume regularization,
we give a low-temperature expansion of such quantities in terms of
the connected diagonal matrix elements, for which we provide explicit
formulas. For special values of the exponent, computations by other
methods are available and used to validate our findings. Our results
can also be interpreted as a further support for a previous conjecture
about the connection between finite- and infinite-volume form factors
valid up to terms exponentially decaying in the volume.
\end{abstract}

\section{Introduction}

Correlation functions and expectation values of operators are important
objects in quantum field theory, both from the theoretical and phenomenological
point of view. Integrable quantum field theories have numerous applications
to condensed matter systems; given that experiments are necessarily
conducted at nonzero temperature the construction of finite temperature
expectation values and correlation functions in integrable quantum
field theories is an interesting problem. Almost fifteen years ago,
LeClair and Mussardo \cite{Leclair:1999ys} put forward a conjecture
for both the one-point and the two-point functions of integrable models
with diagonal scattering, expressed as a spectral series using exact
form factors and the thermodynamic Bethe ansatz. In \cite{Lukyanov:2000jp},
another approach to finite temperature expectation values in the sine-Gordon
model was proposed by Lukyanov, and more recently Negro and Smirnov
provided a resummation of the spectral series of the one-point functions,
again in the sinh-Gordon model \cite{Negro:2013wga}. 

For generic one-point functions, the LeClair-Mussardo proposal was
eventually proven to be valid in \cite{Pozsgay:2010xd}, using the
finite volume form factor formalism introduced in \cite{Pozsgay:2007kn,Pozsgay:2007gx}.
Concerning two-point functions, their proposal is more controversial
\cite{Saleur:1999hq} and probably not valid in its original form.
However, the finite volume form factor formalism provides an alternative
and systematic method to construct the two-point function. This approach
solves the problem faced by earlier studies which could not resolve
the issues related to a proper regularization of kinematical singularities
of the form factors \cite{Konik:2001gf,Altshuler:2005ty}. An early
implementation of the finite volume approach for the two-point functions
was used to describe finite temperature line shapes and dynamical
correlations \cite{Essler:2007jp,Essler:2009zz}. The full formalism
itself was developed in \cite{Pozsgay:2010cr,Szecsenyi:2012jq}. We
note that an alternative approach to thermal correlations was developed
by Doyon \cite{Doyon:2006pv,Chen:2013mda}, however, at present it
seems to be confined to the Ising model.

The finite volume form factor methods were recently shown to yield
results agreeing with other approaches in non-equilibrium steady state
systems \cite{Castro-Alvaredo:2013yla}, and are also relevant in
the context of quantum quenches \cite{Schuricht:2012kr,Mussardo:2013cea,Sotiriadis:2013fca}. 

Presently, the available results on form factor expansions of thermal
correlators in integrable field theory are limited to the case of
diagonal scattering. Conversely, much less is known about non-diagonal
integrable field theories: this is partly due to the fact that the
LeClair-Mussardo expansion, in its original formulation, requires
the solution of the thermodynamic Bethe ansatz equations, which are
considerably more difficult \cite{Balog:2003xd} when the theory is
not diagonal. The finite volume formalism independently provides a
way to extend the results to non-diagonal scattering, and recently
finite volume form factors for non-diagonal scattering were constructed
\cite{Feher:2011fn,Palmai:2012kf}. Albeit at present diagonal matrix
elements of multi-soliton states are still not completely known, the
available results permit the evaluation of the spectral series below
the three-particle threshold. In this paper, we take the first step
and consider finite-temperature expectation values in the sine-Gordon
model, i.e. the one-point functions, which is performed in section
\ref{sec:One-point-functions-at}. In section \ref{sec:Connected-diagonal-form}
we construct the connected diagonal matrix elements of the exponential
operators, which allows the evaluation of the series for these observables.
Exponential operators are particularly useful because they appear
in many physical systems in one dimension, in connection with the
characterization of lattice models at low temperatures, by passing
to a continuous description through an effective bosonic action (see,
e.g., \cite{giamarchi2004quantum,A88,EK04}). In addition, exponential
operators generate all the normal-ordered powers of the sine-Gordon
field, provided it is possible to compute their expectation values
with generic parameter in the exponent.

In section \ref{sec:The-trace-of} we compare the spectral series
for the case of the trace of the energy momentum tensor to results
that follow from the NLIE approach \cite{Klumper:1991b,Destri:1992qk,Destri:1994bv,Feverati:1998dt}
and find very good agreement. Unfortunately, for reasons discussed
towards the end of section \ref{sec:The-trace-of}, we cannot perform
an analogous numerical verification of our method for other operators
at present. Nevertheless, our present results provide a nontrivial
verification of the method and an analytic check of the form of the
diagonal matrix elements conjectured in \cite{Palmai:2012kf} (where
this conjecture was tested numerically against TCSA).

\section{\label{sec:One-point-functions-at}One-point functions at finite
temperature}

The classical action of the sine-Gordon (SG) field theory is:

\begin{equation}
\mathcal{S}=\int_{-\infty}^{\infty}dt\int_{-\infty}^{\infty}dx\left[\frac{1}{2}\partial_{\nu}\phi\partial^{\nu}\phi+\lambda\cos\beta\phi\right]\label{eq:SGaction}
\end{equation}
where ${{\lambda}}$ and $\beta$ are real parameters,
of which $\beta$ is dimensionless and $\lambda$ determines the mass
scale of the model. Classically $\lambda$ has dimension of mass squared,
but in the quantum theory it acquires an anomalous dimension
\[
\lambda\propto[mass]^{2-\beta^{2}/4\pi}
\]
 The fundamental excitations of the model are known to be the soliton,
with mass $m$ and unit topological charge, and the antisoliton, with
equal mass and opposite charge; the exact relation between $\lambda$
and $m$ has been derived by Zamolodchikov in \cite{Zamolodchikov:1995xk}. 

In addition to solitons, the spectrum may also contain breathers which
are bound states of a soliton and antisoliton; after quantization
their spectrum becomes discrete and only a finite number of such states
exists. Introducing the parameter $\xi=\frac{\beta^{2}}{8\pi-\beta^{2}}$,
it is possible to distinguish two regimes: a repulsive one $\xi>1$,
in which only the soliton and the antisoliton are present in the spectrum,
and an attractive one $\xi<1$, in which $\left\lfloor 1/\xi\right\rfloor $
different bound states (breathers), whose mass is
\begin{equation}
m_{b}=2m\sin\frac{\pi\xi b}{2}\;,\quad1\le b<\left\lfloor \frac{1}{\xi}\right\rfloor ,\label{eq:BreatherMass}
\end{equation}
 are allowed.

The scattering matrix between the elementary excitations of the system
has been computed in \cite{zam-zam}; the non-zero elements of the
$S$-matrix in the solitonic sector are

\begin{eqnarray}
S_{sa}^{sa}(\theta)=S_{as}^{as}(\theta) & = & S_{0}(\theta)\frac{\sinh\frac{\theta}{\xi}}{\sinh\frac{i\pi-\theta}{\xi}}\nonumber \\
S_{as}^{sa}(\theta)=S_{sa}^{as}(\theta) & = & S_{0}(\theta)\frac{\sinh\frac{i\pi}{\xi}}{\sinh\frac{i\pi-\theta}{\xi}}\nonumber \\
S_{ss}^{ss}(\theta)=S_{aa}^{aa}(\theta) & = & S_{0}(\theta)\label{eq:Smatrix}
\end{eqnarray}
where
\begin{eqnarray}
S_{0}(\theta) & = & -\exp\left\{ -i\:\intop_{0}^{\infty}dt\frac{\sinh\frac{\pi(1-\xi)t}{2}\sin\theta t}{t\sinh\frac{\pi\xi t}{2}\cosh\frac{\pi t}{2}}\right\} \label{eq:S0}
\end{eqnarray}
The $S$-matrix elements involving breathers are diagonal and can
also be found in \cite{zam-zam}.

Continued to Euclidean time $\tau=-it$, the action 
\begin{equation}
\mathcal{S}_{E}=\int_{0}^{R}d\tau\int_{-\infty}^{\infty}dx\left[\frac{1}{2}\partial_{\tau}\phi\partial_{\tau}\phi+\frac{1}{2}\partial_{x}\phi\partial_{x}\phi-\lambda\cos\beta\phi\right]
\end{equation}
with periodic boundary conditions in $\tau$ describes the model at
finite temperature $T=1/R$. Note that by swapping the role of Euclidean
time and coordinate, the finite temperature/infinite volume action
can also be considered to be a zero temperature/finite volume action,
and so the one-point functions constructed below also have this dual
physical interpretation.

The exponential fields $e^{ik\beta\phi}$ is the most interesting
class of operators to be studied, both because they serve as a generating
function for all the normal-ordered powers of the SG field and in
connection with one-dimensional lattice systems, where they commonly
emerge as a counterpart of lattice operator via bosonization of the
effective low-temperature field theory. For the case $k=\pm1$ the
expectation value of the exponential operator is identical to that
of the perturbing operator $\cos\beta\phi$, which is in turn related
to the trace of the stress-energy $\Theta$ tensor through \cite{Zamolodchikov:1989cf}
\begin{equation}
\left\langle \Theta\right\rangle =4\pi\lambda(1-\Delta)\left\langle e^{\pm i\beta\phi}\right\rangle \label{eq:ThetaVEV}
\end{equation}
where$\Delta=\beta^{2}/(8\pi)$ is the scaling dimension of the exponential
field at the conformal point. 

The finite temperature one-point function of the exponential operators
is defined by Gibbs average:
\begin{eqnarray}
\left\langle e^{ik\beta\phi}\right\rangle  & = & \frac{\mbox{Tr }e^{-RH}e^{ia\beta\phi}}{\mbox{Tr }e^{-RH}}=\frac{{\displaystyle \sum_{n}}e^{-RE_{n}}\langle n|e^{ia\beta\phi}|n\rangle}{{\displaystyle \sum_{n}}e^{-RH}}\label{eq:gibbs_average}
\end{eqnarray}
where
\begin{equation}
H=\int_{-\infty}^{\infty}dx\left[\frac{1}{2}\partial_{t}\phi\partial_{t}\phi+\frac{1}{2}\partial_{x}\phi\partial_{x}\phi-\lambda\cos\left(\beta\phi\right)\right]
\end{equation}
is the Hamiltonian, and the summation runs over a complete set of
energy eigenstates $|n\rangle$ with energies $E_{n}$.

In infinite volume, the form factors 
\begin{equation}
F_{a_{1}\dots a_{n}}^{\mathcal{O}}(\theta_{1},\dots,\theta_{N})=\langle0|\mathcal{O}|\theta_{1},\ldots,\theta_{N}\rangle_{a_{1}\dots a_{n}}
\end{equation}
of local operators can be computed exactly using the form factor bootstrap
\cite{KW78,BKW78,Kirillov:1987jp}, from which any multi-particle
matrix element can be reconstructed using crossing symmetry. Here
\[
|\theta_{1},\ldots,\theta_{N}\rangle_{a_{1}\dots a_{n}}
\]
denotes a multi-particle state composed of particles with species
$a_{1},\dots,a_{N}$ and rapidities $\theta_{1},\dots,\theta_{N}$.
The analytic structure of the form factors is fixed by a set of equations,
which are built upon the factorized scattering of the model as input.
Local operators of a given model can be defined as towers of solutions
of the form factor bootstrap equations \cite{smirnov1992form}. For
many integrable models, including sine-Gordon theory, the exact solutions
are known \cite{smirnov1992form,Lukyanov:1997bp,Babujian:1998uw},
therefore the spectral sum (\ref{eq:gibbs_average}) can be evaluated
in principle.

However, due to the singularity structure which arises from the form
factor axioms, the diagonal matrix elements of the fields are not
well-defined, hence the spectral sum needs to be regularized. The
regularization of form factors by using a finite volume setting is
a useful technique for dealing with low-temperature expansions of
one-point and two-point functions \cite{Pozsgay:2007gx,Pozsgay:2010cr,Szecsenyi:2012jq}.
To evaluate the one-point function one can apply the method outlined
in \cite{Pozsgay:2007gx}; however, a careful extension of the approach
is necessary due to the presence of non-diagonal scattering, which
can be performed using the recent results in \cite{Feher:2011fn,Palmai:2012kf}
on finite volume form factors for non-diagonal scattering. 

We recall that in finite volume the rapidities are quantized and the
space of multi-particle states can be labeled by momentum quantum
numbers $I_{1},\dots,I_{N}$. We introduce the following notation
for them: 
\begin{equation}
|\{I_{1},I_{2},\dots,I_{N}\}\rangle_{L}^{(r)}\label{eq:finvolstate}
\end{equation}
where the index $r$ enumerates the eigenvectors of the $n$-particle
transfer matrix, which can be written as 
\begin{equation}
\mathcal{\mathcal{T}}\left(\theta|\left\{ \theta_{1},\dots,\theta_{N}\right\} \right)_{i_{1}\dots i_{N}}^{j_{1}\dots j_{N}}=\mathcal{S}_{ai_{1}}^{c_{1}j_{1}}(\theta-\theta_{1})\mathcal{S}_{c_{1}i_{2}}^{c_{2}j_{2}}(\theta-\theta_{2})\dots\mathcal{S}_{c_{N-1}i_{N}}^{aj_{N}}(\theta-\theta_{N})
\end{equation}
where $\theta_{1},\dots,\theta_{N}$ are particle rapidities. Due
to factorized scattering, the transfer matrix can be diagonalized
simultaneously for all values of $\theta$:
\begin{equation}
\mathcal{\mathcal{T}}\left(\theta|\left\{ \theta_{1},\dots,\theta_{N}\right\} \right)_{i_{1}\dots i_{N}}^{j_{1}\dots j_{N}}\Psi_{j_{1}\dots j_{n}}^{(r)}\left(\left\{ \theta_{k}\right\} \right)=t^{(r)}\left(\theta,\left\{ \theta_{k}\right\} \right)\Psi_{i_{1}\dots i_{n}}^{(r)}\left(\left\{ \theta_{k}\right\} \right)
\end{equation}
We can assume that the wave function amplitudes $\Psi^{(r)}$ are
normalized and form a complete basis:
\begin{align}
\sum_{i_{1}\dots i_{N}}\Psi_{i_{1}\dots i_{N}}^{(r)}\left(\left\{ \theta_{k}\right\} \right)\Psi_{i_{1}\dots i_{N}}^{(s)}\left(\left\{ \theta_{k}\right\} \right)^{*} & =\delta_{rs}\label{eq:polarization_normalization}\\
\sum_{r}\Psi_{i_{1}\dots i_{N}}^{(r)}\left(\left\{ \theta_{k}\right\} \right)\Psi_{j_{1}\dots j_{N}}^{(r)}\left(\left\{ \theta_{k}\right\} \right)^{*} & =\delta_{i_{1}j_{1}}\dots\delta_{i_{N}j_{N}}\nonumber 
\end{align}
these eigenfunctions describe the possible polarizations of the $N$
particle state with rapidities $\theta_{1},\dots,\theta_{N}$ in the
state indexed by the species quantum numbers $i_{1}\dots i_{N}$.
The transfer matrix can be diagonalized using the algebraic Bethe
ansatz (cf. Appendix A of \cite{Feher:2011aa}), which enables one
to compute the exact form of eigenvalues $t^{(r)}$ and eigenvectors
$\Psi^{(r)}$. The rapidities of the particles in the state (\ref{eq:finvolstate})
can be determined by solving the quantization conditions
\begin{multline}
Q_{j}^{(r)}(\theta_{1},\dots,\theta_{N})=m_{j}L\sinh\theta_{j}+\delta_{j}^{(r)}(\theta_{1},\dots,\theta_{N})=2\pi I_{j},\quad j=1,\dots,N\\
\delta_{j}^{(r)}(\theta_{1},\dots,\theta_{N})=-i\log\left[t^{(r)}\left(\theta_{j},\left\{ \theta_{k}\right\} \right)\right]\label{eq:betheyang_general}
\end{multline}
When considering $N$ rapidities which solve these equations with
given quantum numbers $I_{1},\dots I_{N}$ and a specific polarization
state $r$, they will be written with a tilde as $\tilde{\theta}_{1}^{(N)},\dots,\tilde{\theta}_{N}^{(N)}$.

For states containing up to two particles, the only subspace in which
the transfer matrix has to be diagonalized is the two-dimensional
subspace of states containing one soliton and one antisoliton. The
basis of the eigenstates is given by \cite{Feher:2011fn}
\begin{equation}
\Psi_{ss}=|A_{s}A_{s}\rangle\;,\;\Psi_{aa}=|A_{a}A_{a}\rangle\;,\;\Psi_{+}=\frac{1}{\sqrt{2}}\left(|A_{s}A_{a}\rangle+|A_{a}A_{s}\rangle\right)\;,\;\Psi_{-}=\frac{1}{\sqrt{2}}\left(|A_{s}A_{a}\rangle-|A_{a}A_{s}\rangle\right)
\end{equation}
Assuming that the finite volume energy eigenstates are chosen orthonormal,
the partition function up to (and including) two-particle contribution
expands as

\begin{eqnarray}
Z & = & 1+\sum_{j=s,a}\sum_{\tilde{\theta}}e^{-mR\cosh\tilde{\theta}}+\sum_{b}\sum_{\tilde{\theta}}e^{-m_{b}R\cosh\tilde{\theta}}+\frac{1}{2}\sum_{jj=ss,aa,+,-}\sum_{\tilde{\theta}_{1}^{(2)},\tilde{\theta}_{2}^{(2)}}'e^{-mR\cosh\tilde{\theta}_{1}^{(2)}-mR\cosh\tilde{\theta}_{2}^{(2)}}\nonumber \\
 &  & +\frac{1}{2}\sum_{b_{1}b_{2}}\sum_{\tilde{\theta}_{1}^{(2)}\tilde{\theta}_{2}^{(2)}}'e^{-m_{b_{1}}R\cosh\tilde{\theta}_{1}^{(2)}-m_{b_{2}}R\cosh\tilde{\theta}_{2}^{(2)}}+\sum_{j,b}\sum_{\tilde{\theta}_{1}^{(2)}\tilde{\theta}_{2}^{(2)}}'e^{-mR\cosh\tilde{\theta}_{1}^{(2)}-m_{b}R\cosh\tilde{\theta}_{2}^{(2)}}+\dots
\end{eqnarray}
in which tildes denote rapidities which are quantized according to
the Bethe-Yang equations in finite volume $L$, the index $j=s,a$
is used to denote the elementary solitonic excitations, and the index
$b$ enumerates the breathers. The prime in the summations is a reminder
that states with equal rapidities for the same kind of particle are
not allowed solutions%
\footnote{This is due to the general property (which holds for all 2-dimensional
massive models except free bosons) that for identical particles the
scattering phase is $-1$ when the rapidities coincide, which leads
to a vanishing wave function amplitude.%
} of (\ref{eq:betheyang_general}) and are thus excluded. Furthermore,
the indexes $+,-$ denote the symmetric (antisymmetric) combination
of the neutral soliton-antisoliton states:
\begin{equation}
|\theta_{1},\theta_{2}\rangle_{\pm}=\frac{1}{\sqrt{2}}\left(|\theta_{1},\theta_{2}\rangle_{sa}\pm|\theta_{1},\theta_{2}\rangle_{as}\right)
\end{equation}
The finite temperature expectation value can then be written as 
\begin{equation}
\langle\mathcal{O}\rangle_{R}=\frac{{\displaystyle \sum_{n}}e^{-RE_{n}}\langle n|\mathcal{O}|n\rangle}{Z}
\end{equation}
Following the derivation detailed in \cite{Pozsgay:2007gx}, this
can be expanded as 

\begin{eqnarray}
\frac{\langle\mathcal{O}\rangle_{R}}{Z} & = & \langle\mathcal{O}\rangle+\sum_{j=s,a}\sum_{\tilde{\theta}}e^{-mR\cosh\tilde{\theta}}(_{j}\langle\tilde{\theta}|\mathcal{O}|\tilde{\theta}\rangle_{j}-\langle\mathcal{O}\rangle)+\sum_{b}\sum_{\tilde{\theta}}e^{-m_{b}R\cosh\tilde{\theta}}(_{b}\langle\tilde{\theta}|\mathcal{O}|\tilde{\theta}\rangle_{b}-\langle\mathcal{O}\rangle)\nonumber \\
 &  & +\frac{1}{2}\sum_{jj=ss,aa,+,-}\sum_{\tilde{\theta}_{1}^{(2)}\tilde{\theta}_{2}^{(2)}}e^{-mR\cosh\tilde{\theta}_{1}^{(2)}-mR\cosh\tilde{\theta}_{2}^{(2)}}(\,_{jj}\langle\tilde{\theta}_{1}^{(2)}\tilde{\theta}_{2}^{(2)}|\mathcal{O}|\tilde{\theta}_{2}^{(2)}\tilde{\theta}_{1}^{(2)}\rangle_{jj}-\langle\mathcal{O}\rangle)\nonumber \\
 &  & +\frac{1}{2}\sum_{b_{1}b_{2}}\sum_{\tilde{\theta}_{1}^{(2)}\tilde{\theta}_{2}^{(2)}}e^{-m_{b_{1}}R\cosh\tilde{\theta}_{1}^{(2)}-m_{b_{2}}R\cosh\tilde{\theta}_{2}^{(2)}}(\,_{b_{1}b_{2}}\langle\tilde{\theta}_{1}^{(2)}\tilde{\theta}_{2}^{(2)}|\mathcal{O}|\tilde{\theta}_{2}^{(2)}\tilde{\theta}_{1}^{(2)}\rangle_{b_{2}b_{1}}-\langle\mathcal{O}\rangle)\nonumber \\
 &  & -\frac{1}{2}\sum_{jj=ss,aa}\sum_{\tilde{\theta}_{1}=\tilde{\theta}_{2}=\tilde{\theta}}e^{-2mR\cosh\tilde{\theta}}(\,_{jj}\langle\tilde{\theta}\tilde{\theta}|\mathcal{O}|\tilde{\theta}\tilde{\theta}\rangle_{jj}-\langle\mathcal{O}\rangle)\nonumber \\
 &  & -\frac{1}{2}\sum_{b}\sum_{\tilde{\theta}_{1}=\tilde{\theta}_{2}=\tilde{\theta}}e^{-2m_{b}R\cosh\tilde{\theta}}(\,_{bb}\langle\tilde{\theta}\tilde{\theta}|\mathcal{O}|\tilde{\theta}\tilde{\theta}\rangle_{bb}-\langle\mathcal{O}\rangle)\nonumber \\
 &  & -\frac{1}{2}\sum_{j,k=s,a}\sum_{\tilde{\theta}_{1}^{(1)}\tilde{\theta}_{2}^{(1)}}e^{-mR\cosh\tilde{\theta}_{1}^{(1)}-mR\cosh\tilde{\theta}_{2}^{(1)}}(_{j}\langle\tilde{\theta}_{1}^{(1)}|\mathcal{O}|\tilde{\theta}_{1}^{(1)}\rangle_{j}+_{k}\langle\tilde{\theta}_{2}^{(1)}|O|\tilde{\theta}_{2}^{(1)}\rangle_{k}-2\langle\mathcal{O}\rangle)\nonumber \\
 &  & -\sum_{j,b}\sum_{\tilde{\theta}_{1}^{(2)},\tilde{\theta}_{2}^{(2)}}'e^{-mR\cosh\tilde{\theta}_{1}^{(2)}-m_{b}R\cosh\tilde{\theta}_{2}^{(2)}}(\,_{j}\langle\tilde{\theta}_{1}^{(2)}|\mathcal{O}|\tilde{\theta}_{1}^{(2)}\rangle_{j}+{}_{b}\langle\tilde{\theta}_{2}^{(2)}|O|\tilde{\theta}_{2}^{(2)}\rangle_{b}-2\langle\mathcal{O}\rangle)
\end{eqnarray}
again up to (and including) two-particle contributions. We emphasize
that in the above expression we have explicitly subtracted the terms
in which two elementary excitations with the same topological charge
or two breathers have the same rapidity.

Next we use the relation between the finite and infinite volume form
factors (valid up to exponential terms) conjectured in \cite{Feher:2011fn,Palmai:2012kf}.
The densities of states in rapidity space, corresponding to (\ref{eq:betheyang_general})
are 
\begin{equation}
\rho^{(r)}(\theta_{1},\theta_{2})=\det\left(\begin{array}{cc}
\frac{\partial Q_{1}^{(r)}}{\partial\theta_{1}} & \frac{\partial Q_{1}^{(r)}}{\partial\theta_{2}}\\
\frac{\partial Q_{2}^{(r)}}{\partial\theta_{1}} & \frac{\partial Q_{2}^{(r)}}{\partial\theta_{2}}
\end{array}\right)\qquad r=ss,aa,+,-
\end{equation}
and the finite-volume matrix elements are given by \cite{Feher:2011fn,Palmai:2012kf}
\begin{eqnarray}
\rho^{(jj)}(\tilde{\theta}_{1},\tilde{\theta}_{2})\left(_{jj}\langle\left\{ I_{1}I_{2}\right\} |\mathcal{O}|\left\{ I_{1}I_{2}\right\} \rangle_{jj}-\mathcal{G}_{k}\right) & = & \mathcal{F}_{jj}^{\mathcal{O}(s)}(\tilde{\theta}_{1},\tilde{\theta}_{2})+\rho_{j}(\tilde{\theta}_{1})\mathcal{F}_{j}^{\mathcal{O}}+\rho_{j}(\tilde{\theta}_{2})\mathcal{F}_{j}^{\mathcal{O}}\qquad j=s,a\nonumber \\
\rho^{(\pm)}(\tilde{\theta}_{1},\tilde{\theta}_{2})\left(_{\pm}\langle\left\{ I_{1}I_{2}\right\} |\mathcal{O}|\left\{ I_{1}I_{2}\right\} \rangle_{\pm}-\mathcal{G}_{k}\right) & = & \mathcal{F}_{\pm}^{\mathcal{O}(s)}(\tilde{\theta}_{1},\tilde{\theta}_{2})+\rho_{s}(\tilde{\theta}_{1})\mathcal{F}_{a}^{\mathcal{O}}+\rho_{a}(\tilde{\theta}_{2})\mathcal{F}_{s}^{\mathcal{O}}
\end{eqnarray}
up to terms that vanish exponentially for large $L$. In the above
formulas, the symmetric evaluation is defined as 
\begin{eqnarray}
\mathcal{F}_{\pm}^{\mathcal{O}(s)}(\theta_{1},\theta_{2}) & = & \lim_{\epsilon\to0}F_{\pm}^{\mathcal{O}}(\theta_{2}+i\pi+\epsilon,\theta_{1}+i\pi+\epsilon,\theta_{1},\theta_{2})\\
\mathcal{F}_{jj}^{\mathcal{O}(s)}(\theta_{1},\theta_{2}) & = & \lim_{\epsilon\to0}F_{\bar{j}\bar{j}jj}^{\mathcal{O}}(\theta_{2}+i\pi+\epsilon,\theta_{1}+i\pi+\epsilon,\theta_{1},\theta_{2})
\end{eqnarray}
where the $(\pm,\pm)$-polarized form factors $F_{\pm}^{\mathcal{O}}$
are defined by 
\begin{eqnarray}
F_{r}^{\mathcal{O}}(\theta_{2}'+i\pi,\theta_{1}'+i\pi,\theta_{1},\theta_{2}) & = & \frac{1}{2}\Bigg[F_{sasa}^{\mathcal{O}}(\theta_{2}'+i\pi,\theta_{1}'+i\pi,\theta_{1},\theta_{2})+rF_{saas}^{\mathcal{O}}(\theta_{2}'+i\pi,\theta_{1}'+i\pi,\theta_{1},\theta_{2})\nonumber \\
 & + & rF_{assa}^{\mathcal{O}}(\theta_{2}'+i\pi,\theta_{1}'+i\pi,\theta_{1},\theta_{2})+F_{asas}^{\mathcal{O}}(\theta_{2}'+i\pi,\theta_{1}'+i\pi,\theta_{1},\theta_{2})\Bigg]
\end{eqnarray}
In addition
\begin{eqnarray}
\rho_{s}(\theta)=\rho_{a}(\theta) & = & mL\cosh\theta\nonumber \\
\mathcal{F}_{s}^{k} & = & F_{sa}^{\mathcal{O}}(\theta+i\pi,\theta)\\
\mathcal{F}_{a}^{k} & = & F_{as}^{\mathcal{O}}(\theta+i\pi,\theta)
\end{eqnarray}
(in fact $\mathcal{F}_{s}^{k}=\mathcal{F}_{a}^{k}$ due to charge
conjugation invariance). Following \cite{Pozsgay:2007gx}, we can
also express these results with the connected part of the diagonal
matrix elements which is defined as follows. Consider the form factor

\begin{equation}
F_{\bar{b}\bar{a}ab}^{\mathcal{O}}\left(\theta_{2}+i\pi+\epsilon_{2},\theta_{1}+i\pi+\epsilon_{1},\theta_{1},\theta_{2}\right)
\end{equation}
in which kinematical (simple) poles appear as the regulators $\epsilon_{1,2}\to0$
independently; the connected form factor $\mathcal{F}_{ab}^{k(c)}(\theta_{1},\theta_{2})$
is defined as the part which is nonsingular in both $\epsilon_{1,2}$.
Using the same arguments as in \cite{Pozsgay:2007gx}, it can be easily
checked that the symmetric and connected diagonal matrix elements
are related by

\begin{eqnarray}
\mathcal{F}_{jj}^{\mathcal{O}(s)}(\theta_{1},\theta_{2}) & = & \mathcal{F}_{jj}^{\mathcal{O}(c)}(\theta_{1},\theta_{2})+4\pi G_{0}(\theta_{21})\mathcal{F}_{j}^{\mathcal{O}}\nonumber \\
(\mathcal{F}_{as}^{\mathcal{O}}+\mathcal{F}_{sa}^{\mathcal{O}})^{(s)}(\theta_{1},\theta_{2}) & = & (\mathcal{F}_{as}^{\mathcal{O}}+\mathcal{F}_{sa}^{\mathcal{O}})^{(c)}(\theta_{1},\theta_{2})-2\pi(G_{1}+\bar{G}_{1})(\theta_{21})(\mathcal{F}_{s}^{\mathcal{O}}+\mathcal{F}_{a}^{\mathcal{O}})
\end{eqnarray}
where the function $G_{0}$ is the logarithmic derivative of the soliton-soliton
scattering phase (\ref{eq:S0}): 
\begin{equation}
G_{0}\left(\theta\right)=\frac{1}{2\pi i}\partial_{\theta}\log S_{0}\left(\theta\right)\label{eq:G0}
\end{equation}
and $G_{1}\left(\theta\right)=G_{0}\left(\theta+i\pi\right)$, while
$\bar{G}_{1}\left(\theta\right)$ is its complex conjugate.

On the other hand, for the states in which the scattering among the
particles is diagonal, such as the breather-breather and the soliton-breather,
the finite volume matrix elements are known from \cite{Pozsgay:2007gx}:
\begin{eqnarray}
\rho^{(b_{1}b_{2})}(\tilde{\theta}_{1},\tilde{\theta}_{2})\left(_{b_{1}b_{2}}\langle\left\{ I_{1}I_{2}\right\} |\mathcal{O}|\left\{ I_{1}I_{2}\right\} \rangle_{b_{1}b_{2}}-\mathcal{G}_{k}\right) & = & \mathcal{F}_{b_{1}b_{2}}^{\mathcal{O}(s)}(\tilde{\theta}_{1},\tilde{\theta}_{2})+\rho_{b_{1}}(\tilde{\theta}_{1})\mathcal{F}_{b_{2}}^{\mathcal{O}}+\rho_{b_{2}}(\tilde{\theta}_{2})\mathcal{F}_{b_{1}}^{\mathcal{O}}\nonumber \\
\rho^{(jb)}(\tilde{\theta}_{1},\tilde{\theta}_{2})\left(_{jb}\langle\left\{ I_{1}I_{2}\right\} |\mathcal{O}|\left\{ I_{1}I_{2}\right\} \rangle_{jb}-\mathcal{G}_{k}\right) & = & \mathcal{F}_{jb}^{\mathcal{O}(s)}(\tilde{\theta}_{1},\tilde{\theta}_{2})+\rho_{j}(\tilde{\theta}_{1})\mathcal{F}_{b}^{\mathcal{O}}+\rho_{b}(\tilde{\theta}_{2})\mathcal{F}_{j}^{\mathcal{O}}
\end{eqnarray}
which can be expressed in terms of the connected form factors as 
\begin{eqnarray}
\mathcal{F}_{b_{1}b_{2}}^{\mathcal{O}(s)}(\theta_{1},\theta_{2}) & = & \mathcal{F}_{b_{1}b_{2}}^{\mathcal{O}(c)}(\theta_{1},\theta_{2})+2\pi G_{b_{1}b_{2}}(\theta_{21})\left(\mathcal{F}_{b_{1}}^{\mathcal{O}}+\mathcal{F}_{b_{2}}^{\mathcal{O}}\right)\nonumber \\
\mathcal{F}_{jb}^{\mathcal{O}(s)}(\theta_{1},\theta_{2}) & = & \mathcal{F}_{jb}^{\mathcal{O}(c)}(\theta_{1},\theta_{2})+2\pi G_{jb}(\theta_{21})\left(\mathcal{F}_{j}^{\mathcal{O}}+\mathcal{F}_{b}^{\mathcal{O}}\right)
\end{eqnarray}
with $j=s,a$ for soliton/antisoliton, and $b$ standing for the breather
kind, where
\begin{equation}
\mathcal{F}_{b}^{\mathcal{O}}=F_{bb}^{\mathcal{O}}(\theta+i\pi,\theta)
\end{equation}
and the function $G_{jb}(\theta)$, is defined analogously to (\ref{eq:G0}),
but starting from the soliton-breather scattering phase 
\begin{eqnarray}
2\pi G_{jb}\left(\theta\right) & = & -\frac{4\cos\frac{b\pi\xi}{2}\cosh\theta}{\cos\left(n\pi\xi\right)+\cosh\left(2\theta\right)}\nonumber \\
 &  & -\sum_{l=1}^{b-1}\left(\tan\frac{\pi(1-(2l-n)\xi)-2i\theta}{4}+\tan\frac{\pi(1-(2l-n)\xi)+2i\theta}{4}\right)
\end{eqnarray}
while in turn, the scattering phase between breathers $b_{1}$ and
$b_{2}$ defines the function:
\begin{eqnarray}
2\pi G_{b_{1}b_{2}}(\theta) & = & \frac{4\cosh\theta\sin\left(\frac{b_{1}+b_{2}}{2}\pi\xi\right)}{\cos(b_{1}+b_{2})\pi\xi-\cosh\left(2\theta\right)}+\frac{4\cosh\theta\sin\left(\frac{b_{1}-b_{2}}{2}\pi\xi\right)}{\cos(b_{1}-b_{2})\pi\xi-\cosh\left(2\theta\right)}\nonumber \\
 & - & \sum_{l=1}^{n-1}\Bigg\{\frac{\sin\frac{(2l+b_{2}-b_{1})\pi\xi}{2}}{\sinh\frac{2\theta-i(2l+b_{2}-b_{1})\pi\xi}{4}\sinh\frac{2\theta+i(2l+b_{2}-b_{1})\pi\xi}{4}}\nonumber \\
 &  & -\frac{\sin\frac{(2l-b_{1}-b_{2})\pi\xi}{2}}{\cosh\frac{2\theta-i(2l-b_{1}-b_{2})\pi\xi}{4}\cosh\frac{2\theta+i(2l-b_{1}-b_{2})\pi\xi}{4}}\Bigg\}
\end{eqnarray}
for $b_{1}\ge b_{2}$.

Substituting into (\ref{eq:expansion}), and keeping terms up to two
particles:
\begin{eqnarray}
\langle\mathcal{O}\rangle^{R} & = & \langle\mathcal{O}\rangle+\sum_{j=s,a}\int\frac{d\theta}{2\pi}e^{-mR\cosh\theta}\mathcal{F}_{j}^{\mathcal{O}}-\sum_{j=s,a}\int\frac{d\theta}{2\pi}e^{-2mR\cosh\theta}\mathcal{F}_{j}^{\mathcal{O}}\nonumber \\
 &  & +\sum_{b}\int\frac{d\theta}{2\pi}e^{-m_{b}R\cosh\theta}\mathcal{F}_{b}^{\mathcal{O}}-\sum_{b}\int\frac{d\theta}{2\pi}e^{-2m_{b}R\cosh\theta}\mathcal{F}_{b}^{\mathcal{O}}\nonumber \\
 &  & +\frac{1}{2}\sum_{j=s,a}\int\frac{d\theta_{1}}{2\pi}\int\frac{d\theta_{2}}{2\pi}e^{-mR\cosh\theta_{1}-mR\cosh\theta_{2}}(\mathcal{F}_{jj}^{\mathcal{O}(c)}(\theta_{21})+2\cdot2\pi G_{0}(\theta_{21})\mathcal{F}_{j}^{\mathcal{O}})\nonumber \\
 &  & +\frac{1}{2}\int\frac{d\theta_{1}}{2\pi}\int\frac{d\theta_{2}}{2\pi}e^{-mR\cosh\theta_{1}-mR\cosh\theta_{2}}\Bigg[\mathcal{F}_{+}^{\mathcal{O}(c)}(\theta_{21})+\mathcal{F}_{-}^{\mathcal{O}(c)}(\theta_{21})\nonumber \\
 &  & -2\pi\left(G_{1}(\theta_{21})+\bar{G}_{1}(\theta_{21})\right)\Bigg]\left(\mathcal{F}_{s}^{\mathcal{O}}+\mathcal{F}_{a}^{\mathcal{O}}\right)\nonumber \\
 &  & +\frac{1}{2}\sum_{b_{1}b_{2}}\int\frac{d\theta_{1}}{2\pi}\int\frac{d\theta_{2}}{2\pi}e^{-m_{b_{1}}R\cosh\theta_{1}-m_{b_{2}}R\cosh\theta_{2}}\left(\mathcal{F}_{b_{1}b_{1}}^{\mathcal{O}(c)}(\theta_{21})+2\pi G_{b_{1}b_{2}}(\theta_{21})(\mathcal{F}_{b_{1}}^{\mathcal{O}}+\mathcal{F}_{b_{2}}^{\mathcal{O}})\right)\nonumber \\
 &  & +\sum_{j,b}\int\frac{d\theta_{1}}{2\pi}\int\frac{d\theta_{2}}{2\pi}e^{-mR\cosh\theta_{1}-m_{b}R\cosh\theta_{2}}\left(\mathcal{F}_{jb}^{\mathcal{O}(c)}(\theta_{21})+2\pi G_{jb}(\theta_{21})(\mathcal{F}_{j}^{\mathcal{O}}+\mathcal{F}_{b}^{\mathcal{O}})\right)+\dots\label{eq:expansion}
\end{eqnarray}
which gives the low-temperature expansion of the expectation value
of the local field $\mathcal{O}$ in the sine-Gordon theory, up to
and including two-particle contributions.

\section{\label{sec:Connected-diagonal-form}Connected diagonal matrix elements
of the exponential fields}

\subsection{Form factors of exponential fields\label{sub:Form-factors-review}}

Multi-soliton form factors of exponential operators in the sine-Gordon
model 
\begin{equation}
F_{a_{1},\ldots,a_{n}}^{k}(\theta_{1},\ldots,\theta_{n})=\langle0|e^{ik\beta\phi}|\theta_{1},\ldots,\theta_{n}\rangle_{a_{1},\ldots,a_{n}}
\end{equation}
were obtained by Lukyanov in \cite{Lukyanov:1997bp} exploiting the
bosonized form of the Zamolodchikov-Faddeev operators \cite{Lukyanov:1993pn}:
\begin{eqnarray}
Z_{s}(\theta) & = & \sqrt{\frac{i\mathcal{C}_{2}}{4\mathcal{C}_{1}}}e^{ik\theta}e^{i\phi(\theta)}\nonumber \\
Z_{a}(\theta) & = & \sqrt{\frac{i\mathcal{C}_{2}}{4\mathcal{C}_{1}}}e^{-ik\theta}\sum_{\sigma=\pm}\sigma e^{\sigma\frac{4\pi^{2}i}{\beta^{2}}}\int_{\Gamma_{\sigma}}\frac{d\gamma}{2\pi}e^{(1-2k-8\pi\beta^{-2})\gamma}W(\sigma(\gamma-\theta)):e^{-i\bar{\phi}(\gamma)}e^{i\phi(\theta)}:\label{eq:FZboson}
\end{eqnarray}
in which $:\;:$ denotes the appropriate normal ordering while $\phi$,
$\bar{\phi}$ are generalized free fields defined in \cite{Lukyanov:1993pn}
and the contours $\Gamma_{\pm}$ are specified below. These operators
satisfy the algebra of asymptotic soliton/antisoliton creation operators

\begin{equation}
Z_{a}(\theta_{1})Z_{b}(\theta_{2})=S_{ba}^{cd}(\theta_{21})Z_{c}(\theta_{2})Z_{d}(\theta_{1})
\end{equation}
with the scattering matrix (\ref{eq:Smatrix}).

The contractions $\langle\langle\quad\rangle\rangle$ of the vertex
operators are defined as follows:

\begin{eqnarray}
\langle\langle e^{i\phi(\theta_{1})}e^{i\phi(\theta_{2})}\rangle\rangle & = & G(\theta_{2}-\theta_{1})\label{eq:defG}\\
\langle\langle e^{i\phi(\theta_{1})}e^{-i\bar{\phi}(\theta_{2})}\rangle\rangle & = & W(\theta_{2}-\theta_{1})=\frac{1}{G(\theta_{2}-\theta_{1}-i\frac{\pi}{2})G(\theta_{2}-\theta_{1}+i\frac{\pi}{2})}\label{eq:defW}\\
\langle\langle e^{-i\bar{\phi}(\theta_{1})}e^{-i\bar{\phi}(\theta_{2})}\rangle\rangle & = & \bar{G}(\theta_{2}-\theta_{1})=\frac{1}{W(\theta_{2}-\theta_{1}-i\frac{\pi}{2})W(\theta_{2}-\theta_{1}+i\frac{\pi}{2})}\label{eq:defGbar}
\end{eqnarray}
where
\begin{eqnarray}
G(\theta) & = & i\mathcal{C}_{1}\sinh\frac{\theta}{2}\exp\left\{ \intop\frac{dt}{t}\frac{\sinh^{2}(1-i\theta/\pi)\sinh(\xi-1)t}{\sinh2t\sinh\xi t\cosh t}\right\} \label{eq:G}\\
W(\theta) & = & \frac{-2}{\cosh\theta}\exp\left\{ -2\intop\frac{dt}{t}\frac{\sinh^{2}(1-i\theta/\pi)\sinh(\xi-1)t}{\sinh2t\sinh\xi t}\right\} =\frac{\hat{W}\left(\theta\right)}{\cosh\theta}\label{eq:W}\\
\bar{G}(\theta) & = & -\frac{\xi\mathcal{C}_{2}}{4}\sinh\theta\sinh\frac{\theta+i\pi}{\xi}\label{eq:Gbar}
\end{eqnarray}
for $\frac{|\xi-1|-\xi}{2}-2<\Im m\,\frac{\theta}{\pi}<\frac{\xi-|\xi-1|}{2}$;
analytic continuations valid for a wider range of rapidities can be
found in \cite{Feher:2011fn,Palmai:2011kb}. 

Using the above definitions, Lukyanov constructed the multi-soliton
form factors in the form
\begin{equation}
\tilde{F}_{a_{1},\ldots,a_{n}}^{k}(\theta_{1},\ldots,\theta_{n})=\mathcal{G}_{k}\langle\langle Z_{a_{1}}(\theta_{1})\dots Z_{a_{n}}(\theta_{n})\rangle\rangle\label{eq:lukyanov_ff}
\end{equation}
where 
\[
\mathcal{G}_{k}=\langle e^{ik\beta\phi}\rangle
\]
is the vacuum expectation value of the field and the integration contours
are specified as follows. Calling ``principal'' poles the singularities
of the function $W(\gamma)$ located at $\gamma=-i\pi/2$, the contours
run above the ``principal'' singularities of the $W$ functions
arising from contraction of a given operator $e^{-i\bar{\phi}}$ with
all fields on its right and below the principal poles originating
from the contraction with the ones on its left. Accordingly, in the
definition (\ref{eq:FZboson}) $\Gamma_{+}$($\Gamma_{-}$) denote
the contours which pass above (below) the pole at $\gamma=\theta+i\pi/2$
($\gamma=\theta-i\pi/2$).

The tilde in (\ref{eq:lukyanov_ff}) refers to the fact that our conventions
for the form factors differ from those of Lukyanov's by the relation
\begin{eqnarray}
F_{a_{1}\dots a_{2n}}^{k}(\theta_{1},\dots,\theta_{2n}) & = & (-1)^{n}\tilde{F}_{-a_{1}\dots-a_{2n}}^{k}(\theta_{1},\dots,\theta_{2n})\nonumber \\
 & = & (-1)^{n}\tilde{F}_{a_{1}\dots a_{2n}}^{-k}(\theta_{2n},\dots,\theta_{1})
\end{eqnarray}
as noted in \cite{Feher:2011fn}. This is due to a difference in the
conventions of the form factor equation, which corresponds to a redefinition
of relative phases of the multi-particle states \cite{Feher:2011fn}. 

Note that all the form factors need to be normalized by the vacuum
expectation value of the exponential field, for which a formula has
been derived in \cite{Lukyanov:1996jj}. It is useful to write that
formula in a way which can be used for any value of $k$. Given two
integers $M,N\ge\max\left(0,\left\lceil \frac{4k\Delta-2-2\left(1-\Delta\right)}{2(1-\Delta)}\right\rceil \right)$
and defining $\Delta=\Delta(\beta)=\beta^{2}/8\pi$, one has:

\begin{eqnarray}
\mathcal{G}_{k} & = & \left(\frac{\sqrt{\pi}\Gamma\left(\frac{1}{2(1-\Delta)}\right)}{2\Gamma\left(\frac{\Delta}{2(1-\Delta)}\right)}\right)^{2k^{2}\Delta}\nonumber \\
 &  & \times\prod_{m=0}^{M}\prod_{n=0}^{N}\left(\frac{\Gamma(1+(1-\Delta)m+n\Delta-2k\Delta)\Gamma(1+(1-\Delta)m+n\Delta+2k\Delta)}{\Gamma(1+(1-\Delta)m+n\Delta)^{2}}\right)^{(-)^{m}}\nonumber \\
 &  & \times\exp\Big\{(-)^{M+1}\sum_{n=0}^{N}\int_{0}^{\infty}\frac{dt}{t}\frac{e^{-(2n+1)\Delta t-2(1-\Delta)(M+1)t}\sinh^{2}(2k\Delta t)}{\sinh t\cosh(1-\Delta)t}\nonumber \\
 &  & \qquad+(-)^{M+1}\int_{0}^{\infty}\frac{dt}{t}\left(\frac{e^{-2(N+1)\Delta t-2(M+1)(1-\Delta)t}\sinh^{2}(2k\Delta t)}{2\sinh t\sinh\Delta t\cosh(1-\Delta)t}-2k^{2}\Delta e^{-2t}\right)\nonumber \\
 &  & \qquad+\sum_{m=0}^{M}(-)^{m}\int_{0}^{\infty}\frac{dt}{t}\left(\frac{e^{-2(N+1)\Delta t-2m(1-\Delta)t}\sinh^{2}(2k\Delta t)}{\sinh t\sinh\Delta t}-4k^{2}\Delta e^{-2t}\right)\Big\}\label{eq:vev-1}
\end{eqnarray}
which can be obtained by exploiting the integral representation of
the logarithm of the Euler's gamma function. Finally, it was noted
in \cite{Feher:2011aa} that to make agreement with TCSA results,
a sign had to be inserted in the diagonal matrix elements with an
odd number of solitons or antisolitons. Therefore, our diagonal matrix
elements can be obtained in the following way:
\begin{equation}
\mathcal{F}_{a_{1}\ldots a_{n}}^{k}\left(\theta_{1},\ldots,\theta_{n}\right)=(-1)^{n}\mathbb{F}_{a_{1}\ldots a_{n}}^{k}\left(\theta_{1},\ldots,\theta_{n}\right)
\end{equation}
where $\mathbb{F}^{k}$ denotes diagonal matrix elements in Lukyanov's
conventions \cite{Lukyanov:1997bp}.

\subsection{The diagonal matrix elements\label{sub:The-diagonal-form}}

As a warm-up and a demonstration of how the analytic continuation
is performed, it is useful to write down the two-particle form factor
in the repulsive regime:
\begin{eqnarray}
\mathcal{F}_{s}^{k} & = & \mathcal{G}_{k}\frac{i\mathcal{C}_{2}}{4\mathcal{C}_{1}}e^{\left(i\pi-\epsilon\right)k}\sum_{\sigma=\pm}\sigma e^{i\frac{\sigma}{2}(1+\frac{1}{\xi})\pi}\intop_{\Gamma_{\sigma}}\frac{d\gamma}{2\pi}\left\langle \left\langle e^{i\phi(\theta+i\pi+\epsilon)}:e^{-i\bar{\phi}(\gamma)}e^{i\phi(\theta)}:\right\rangle \right\rangle W(\sigma(\theta-\gamma))e^{A(\gamma-\theta)}\nonumber \\
 & = & \mathcal{G}_{k}\frac{i\mathcal{C}_{2}}{4}e^{\left(i\pi-\epsilon\right)k}\sum_{\sigma=\pm}\sigma e^{i\frac{\sigma}{2}(1+\frac{1}{\xi})\pi}\Bigg\{-\frac{\hat{W}(-i\frac{\pi}{2})\hat{W}(-i\frac{\pi}{2}-\sigma\epsilon)}{\epsilon}\nonumber \\
 &  & +\intop_{\mathbb{R}}\frac{d\gamma}{2\pi}e^{A\gamma}W(-\sigma\gamma)W(\gamma-i\pi-\epsilon)\Bigg\}\label{eq:Fsa_regularized}
\end{eqnarray}
in which the contour $\Gamma_{+(-)}$ passes above (below) the pole
at $\gamma=i\frac{\pi}{2}$ ($\gamma=-i\frac{\pi}{2}$) and has been
deformed to the real axis in the second line. We also introduced the
notation $A=-2k-\xi^{-1}$. 

Consider now the integral part, which is divergent for $k\ge1/2$.
It can be written as the analytic continuation of a Fourier transform
to imaginary values $z=-iA$:
\begin{eqnarray}
I_{\sigma}(z) & = & \intop_{\mathbb{R}}\frac{d\gamma}{2\pi}e^{iz\gamma}W(-\sigma\gamma)W(\gamma-i\pi)=-\frac{1}{\mathcal{C}_{2}}\intop_{\mathbb{R}}dx\frac{1}{\cosh\frac{\pi(z-x)}{2}}\frac{e^{-\sigma\frac{\pi(1-\xi)x}{2}}}{\cosh\frac{\pi\xi x}{2}}
\end{eqnarray}
where the definition (\ref{eq:defGbar}) has been used. Now the integral
is convergent, but we still need to continue the integral to the value
$z=i\left(2k+1/\xi\right)$ by adding the poles that are crossed in
the contour deformation. The result is then
\begin{eqnarray}
I_{\sigma}\left(i(2k+1/\xi)\right) & =-\frac{1}{\mathcal{C}_{2}} & \Bigg\{4\sum_{m=0}^{\left\lfloor (2k+1/\xi-1)/2\right\rfloor }(-1)^{m}\frac{e^{-\sigma i\frac{\pi(1-\xi)\left(2k+1/\xi-1-2m\right)}{2}}}{\cos\frac{\pi\xi\left(2k+1/\xi-1-2m\right)}{2}}\nonumber \\
 &  & +\intop_{\mathbb{R}}dx\frac{1}{\cosh\frac{\pi(i\left(2k+1/\xi\right)-x)}{2}}\frac{e^{-\sigma\frac{\pi(1-\xi)x}{2}}}{\cosh\frac{\pi\xi x}{2}}\Bigg\}
\end{eqnarray}
and is now convergent for all real values of $k$. Note that this
part is $O(\epsilon^{0})$. The connected part of the matrix
element (\ref{eq:Fsa_regularized}) is the total $O(\epsilon^{0})$
contribution, which can be collected as

\begin{eqnarray}
\mathcal{F}_{s}^{k} & = & \mathcal{G}_{k}\frac{i\mathcal{C}_{2}}{4}e^{i\pi k}\sum_{\sigma=\pm}e^{i\frac{\sigma}{2}(1+\frac{1}{\xi})\pi}\left(\left(1+\left\lfloor \frac{\xi}{2}\right\rfloor \right)\hat{W}(-i\frac{\pi}{2})\hat{W}'(-i\frac{\pi}{2})+\sigma I_{\sigma}\left(i(2k+1/\xi)\right)\right)\label{eq:Fsa_connected}
\end{eqnarray}
This will be compared with the exact formula below in section \ref{sub:Comparison}.
Note that the function $\hat{W}$ defined in (\ref{eq:W}) is regular
in the point $-i\pi/2$ and its derivative can be computed straightforwardly
from the definition.

\begin{figure}
\begin{centering}
\includegraphics[width=0.5\textwidth]{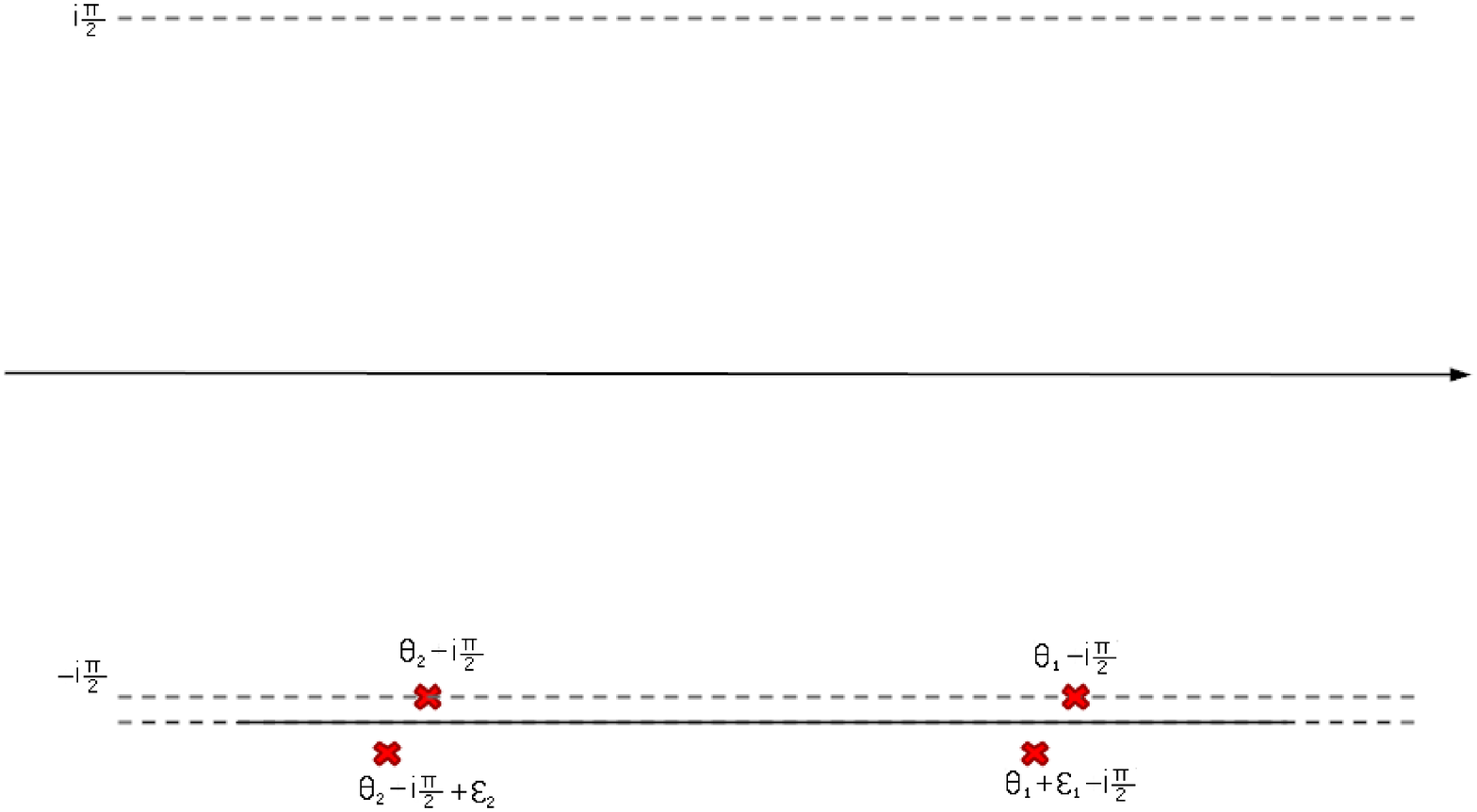}\includegraphics[width=0.5\textwidth]{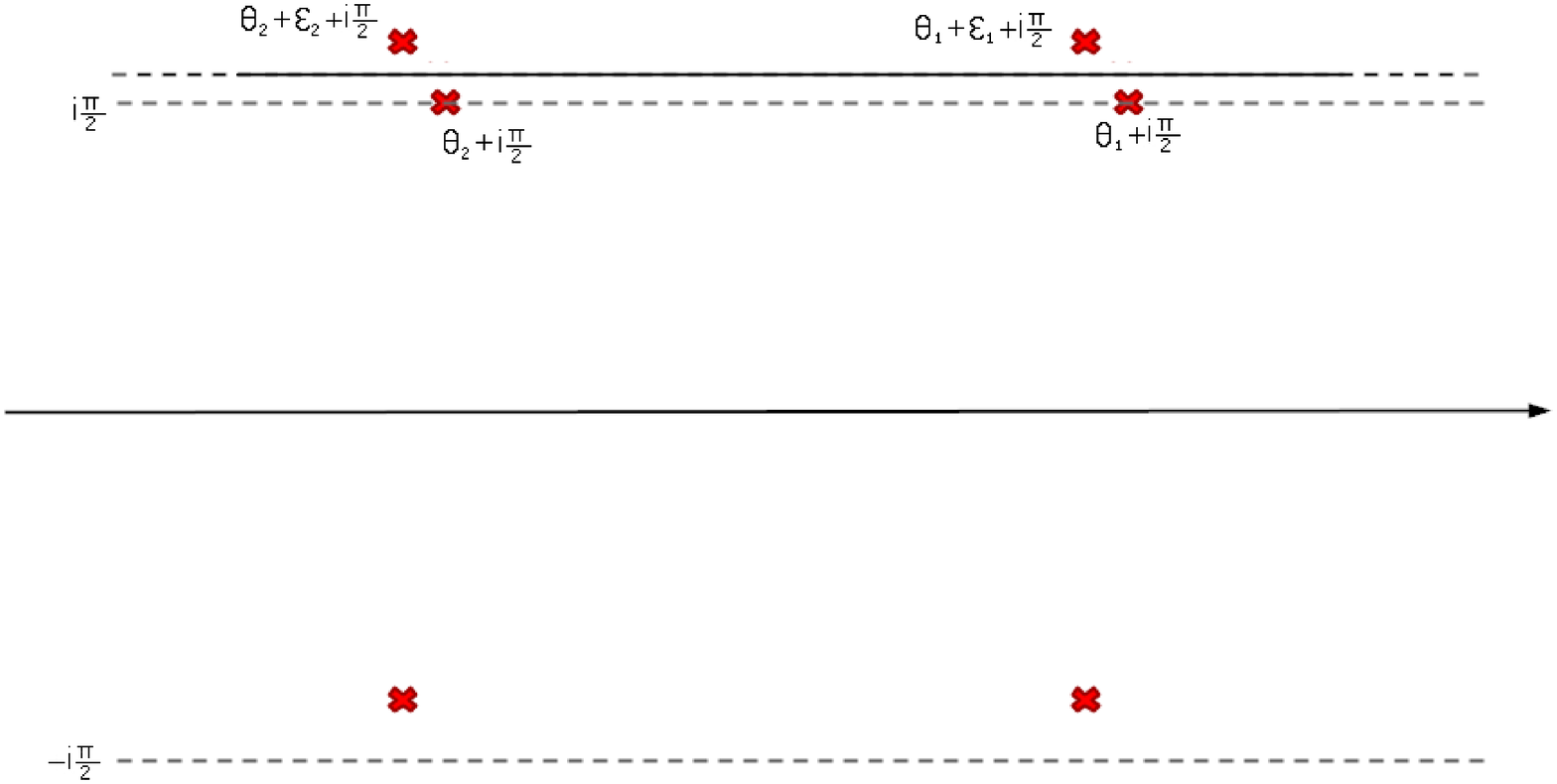}\\
\includegraphics[width=0.5\textwidth]{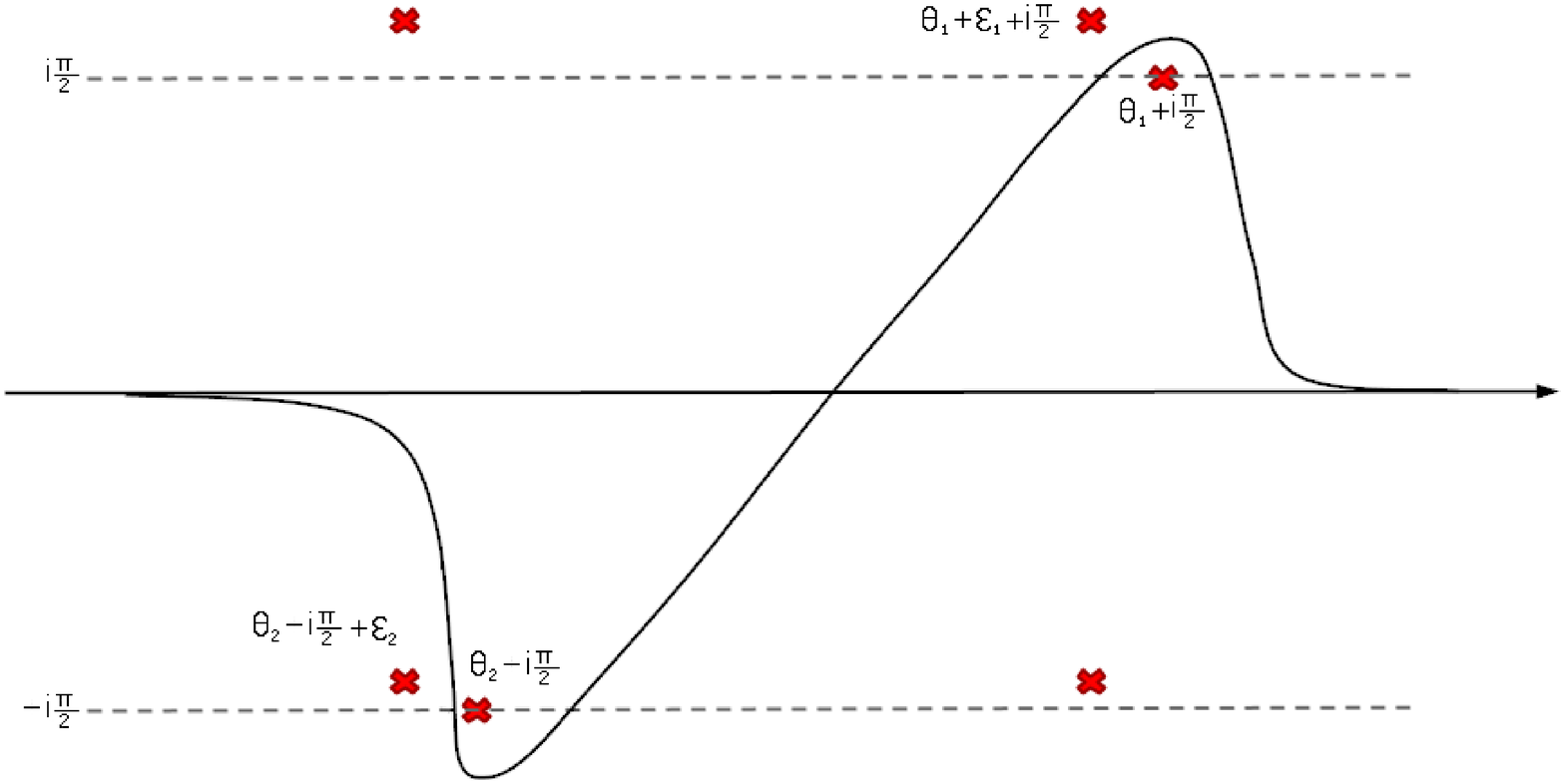}
\par\end{centering}

\caption{Left: Contour $\Gamma_{-}^{(1)}$, right: Contour $\Gamma_{+}^{(2)}$,
below: Contour $\Gamma_{+}^{(1)}\equiv\Gamma_{-}^{(2)}$\label{fig:Contours-ppmm}}
\end{figure}

Let us now write explicitly the regularized diagonal four particle
form factor:
\begin{eqnarray}
 &  & F_{ssaa}^{k}(\theta_{2}+i\pi+\epsilon_{2},\theta_{1}+i\pi+\epsilon_{1},\theta_{1},\theta_{2})=\nonumber \\
 &  & \mathcal{G}_{k}\left(\frac{i\mathcal{C}_{2}}{4\mathcal{C}_{1}}\right)^{2}e^{-k(2\pi i+\epsilon_{1}+\epsilon_{2})}\sum_{\sigma_{1}\sigma_{2}=\pm}\sigma_{1}\sigma_{2}e^{i\frac{\sigma_{1}+\sigma_{2}}{2}(1+\frac{1}{\xi})\pi}\intop_{\Gamma_{\sigma_{1}}^{(1)}}\frac{d\gamma_{1}}{2\pi}W\left(\sigma_{1}(\theta_{1}-\gamma_{1})\right)\nonumber \\
 &  & \times\intop_{\Gamma_{\sigma_{2}}^{(2)}}\frac{d\gamma_{2}}{2\pi}W\left(\sigma_{2}(\theta_{2}-\gamma_{2})\right)e^{A(\gamma_{1}-\theta_{1}+\gamma_{2}-\theta_{2})}\nonumber \\
 &  & \times\left\langle \left\langle e^{i\phi(\theta_{1}+i\pi+\epsilon_{1})}e^{i\phi(\theta_{2}+i\pi+\epsilon_{2})}:e^{-i\bar{\phi}(\gamma_{2})}e^{i\phi(\theta_{2})}::e^{-i\bar{\phi}(\gamma_{1})}e^{i\phi(\theta_{1})}:\right\rangle \right\rangle 
\end{eqnarray}
where we again denote $A=-2k-\xi^{-1}$. The contraction in the last
line contains one factor which is independent of the integrated variables
and reads:
\begin{eqnarray}
\mathcal{A}_{ssaa} & = & \left\langle \left\langle e^{i\phi(\theta_{1}+i\pi+\epsilon_{1})}e^{i\phi(\theta_{2}+i\pi+\epsilon_{2})}e^{i\phi(\theta_{2})}e^{i\phi(\theta_{1})}\right\rangle \right\rangle \nonumber \\
 &  & G\left(\theta_{12}\right)G\left(\theta_{12}-i\pi-\epsilon_{2}\right)G\left(-i\pi-\epsilon_{1}\right)\nonumber \\
 &  & G\left(-i\pi-\epsilon_{2}\right)G\left(\theta_{21}+\epsilon_{21}\right)G\left(\theta_{21}-i\pi-\epsilon_{1}\right)\label{eq:Assaa0}
\end{eqnarray}
with the usual notation $\epsilon_{21}=\epsilon_{2}-\epsilon_{1}$;
on the other hand, the integral parts are as follows
\begin{eqnarray}
 &  & \intop_{\Gamma_{\sigma_{1}}^{(1)}}\frac{d\gamma_{1}}{2\pi}e^{A\gamma_{1}}W(\gamma_{1}+\frac{\theta}{2}-i\pi-\epsilon_{1})W(-\sigma_{1}(\frac{\theta}{2}+\gamma_{1}))W(\gamma_{1}-\frac{\theta}{2}-i\pi-\epsilon_{2})W(\gamma_{1}-\frac{\theta}{2})\\
 &  & \intop_{\Gamma_{\sigma_{2}}^{(2)}}\frac{d\gamma_{2}}{2\pi}e^{A\gamma_{2}}W(\gamma_{2}+\frac{\theta}{2}-i\pi-\epsilon_{1})W(-\frac{\theta}{2}-\gamma_{2})W(\gamma_{2}-\frac{\theta}{2}-i\pi-\epsilon_{2})W(-\sigma_{2}(\gamma_{2}-\frac{\theta}{2}))\bar{G}(\gamma_{21})\nonumber 
\end{eqnarray}
with the contours depicted in figure \ref{fig:Contours-ppmm}.

\begin{figure}
\begin{centering}
\includegraphics[width=0.5\textwidth]{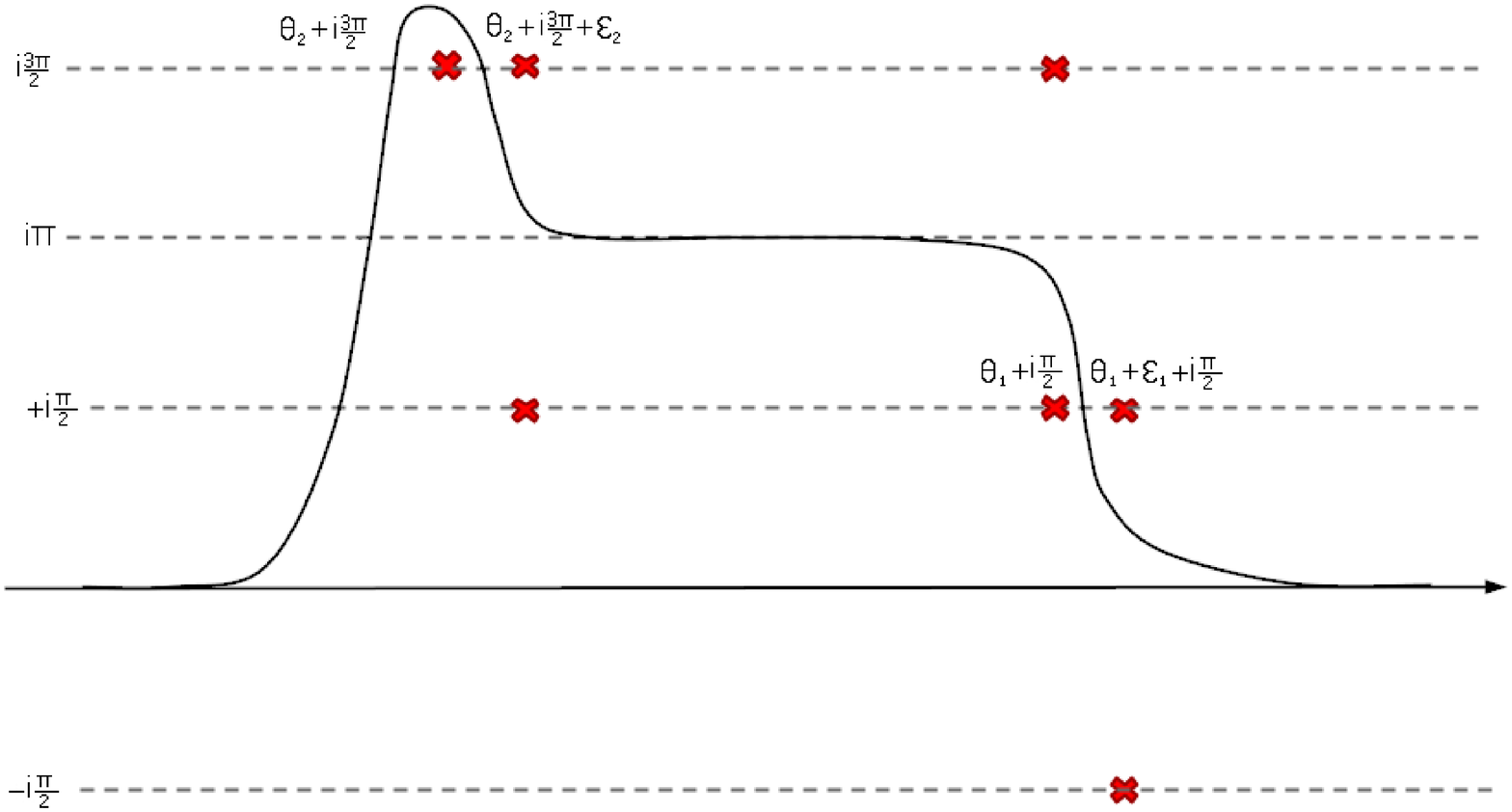}\includegraphics[width=0.5\textwidth]{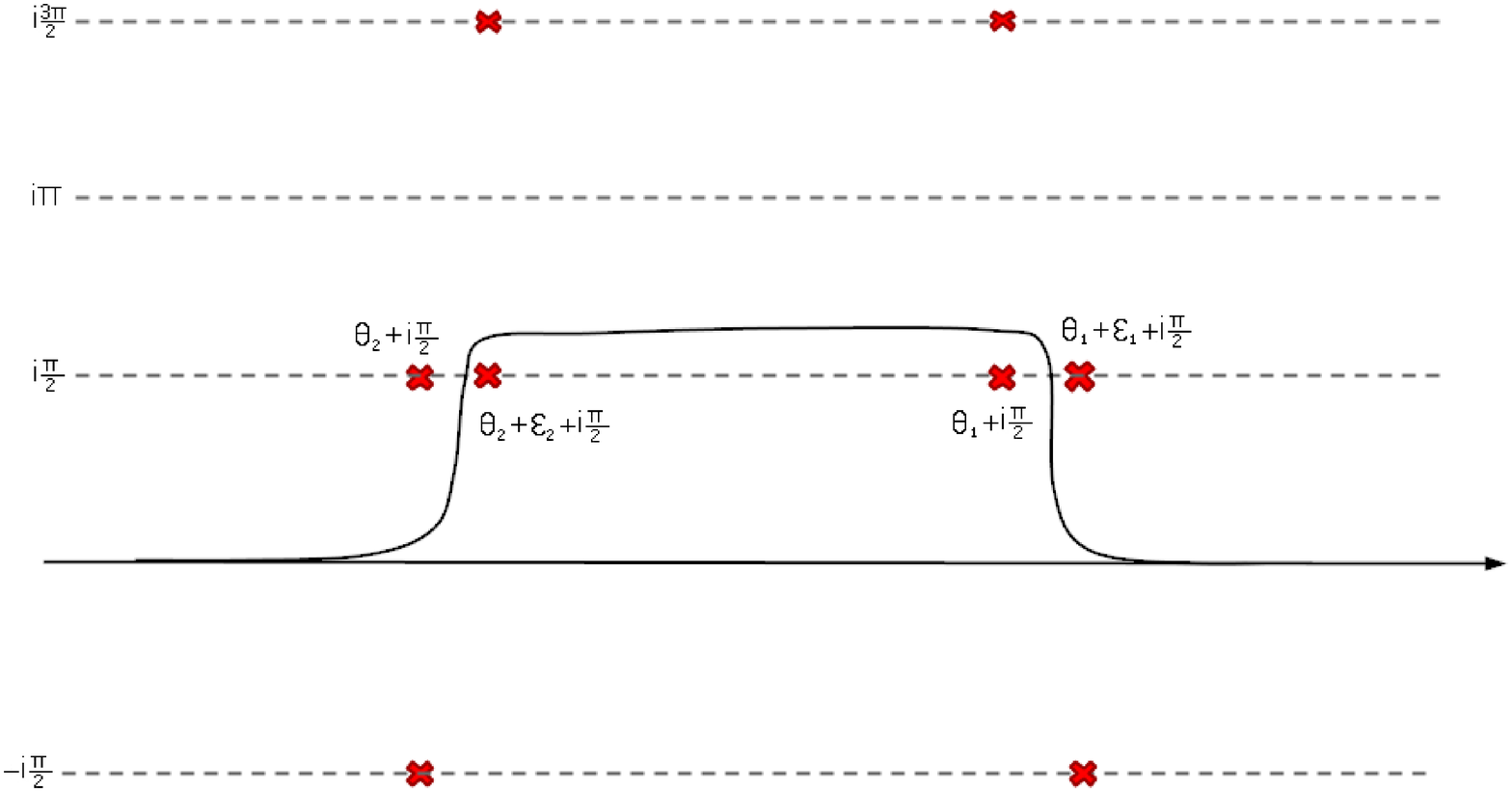}
\par\end{centering}

\caption{Left: Contour $\Gamma_{+}^{(2)}$, Right: Contour $\Gamma_{-}^{(2)}$\label{fig:Contours-pmpm2}}
\end{figure}

\begin{figure}
\begin{centering}
\includegraphics[width=0.5\textwidth]{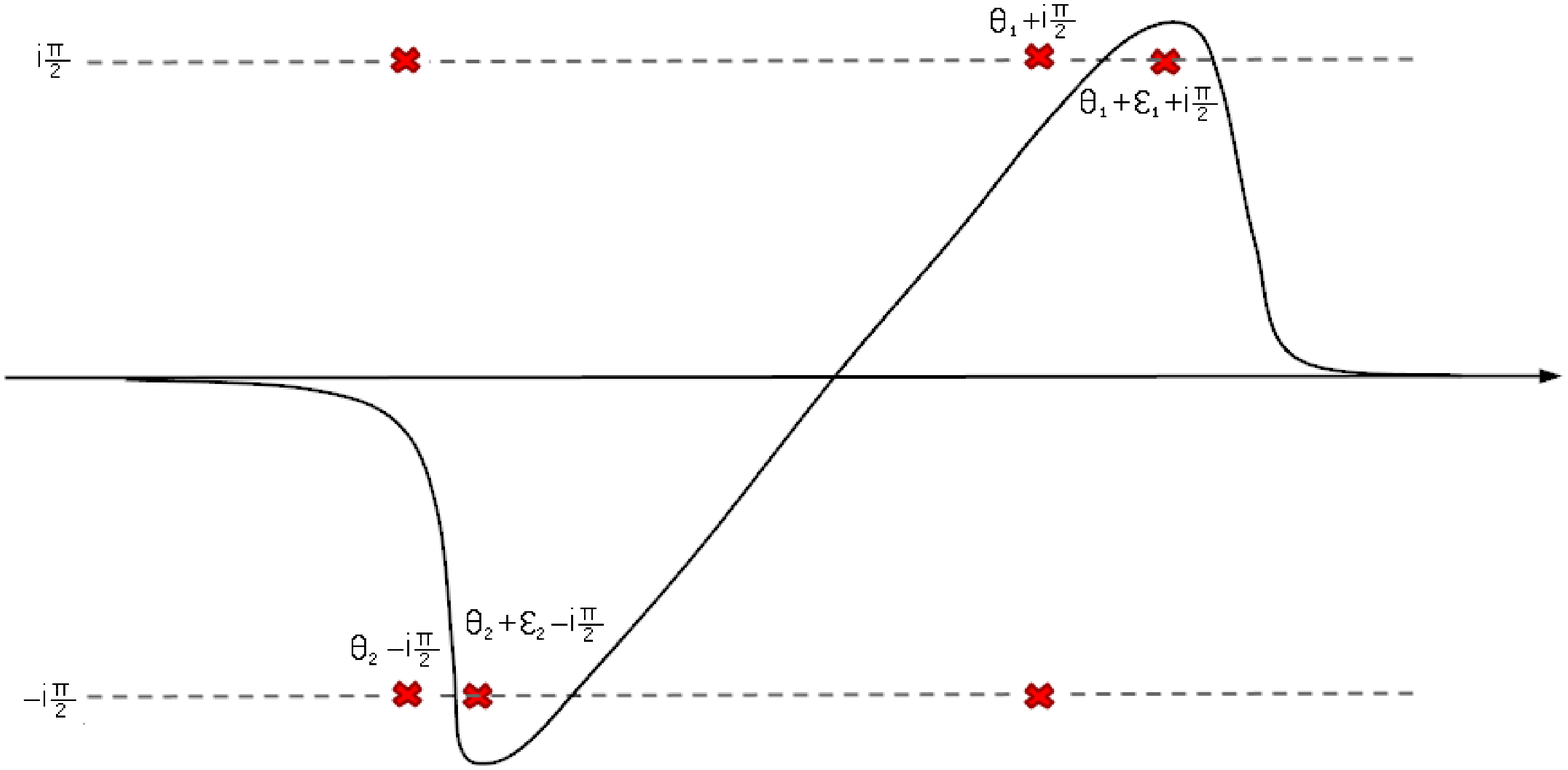}\includegraphics[width=0.5\textwidth]{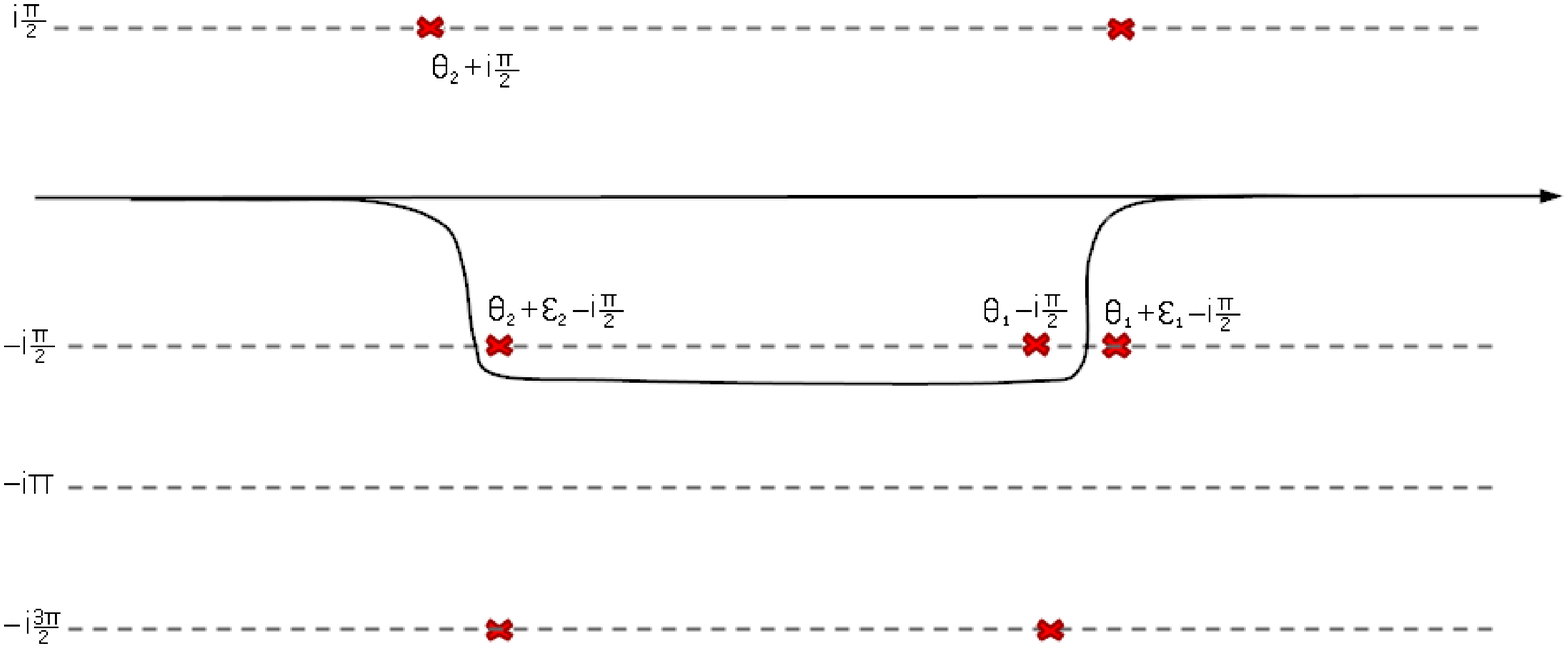}
\par\end{centering}

\caption{Left: Contour $\Gamma_{+}^{(1)}$, Right: Contour $\Gamma_{-}^{(1)}$\label{fig:Contours-pmpm1}}
\end{figure}

Moreover, we have 
\begin{eqnarray}
 &  & F_{sasa}^{k}(\theta_{2}+i\pi+\epsilon_{2},\theta_{1}+i\pi,\theta_{1}+\epsilon_{1},\theta_{2})=\nonumber \\
 &  & \mathcal{G}_{k}\left(\frac{i\mathcal{C}_{2}}{4\mathcal{C}_{1}}\right)^{2}e^{k(\epsilon_{2}-\epsilon_{1})}\sum_{\sigma_{1}\sigma_{2}=\pm}\sigma_{1}\sigma_{2}e^{i\frac{\sigma_{1}+\sigma_{2}}{2}(1+\frac{1}{\xi})\pi}\intop_{\Gamma_{\sigma_{1}}^{(1)}}\frac{d\gamma_{1}}{2\pi}W(\sigma_{1}(\theta_{1}-\gamma_{1}))\\
 &  & \times\intop_{\Gamma_{\sigma_{2}}^{(2)}}\frac{d\gamma_{2}}{2\pi}W(\sigma_{2}(\theta_{2}-\gamma_{2}+i\pi))e^{A(\gamma_{1}-\theta_{1}+\gamma_{2}-\theta_{2}-i\pi)}\nonumber \\
 &  & \times\left\langle \left\langle e^{i\phi(\theta_{1}+i\pi+\epsilon_{1})}:e^{-i\bar{\phi}(\gamma_{2})}e^{i\phi(\theta_{2}+i\pi)}:e^{i\phi(\theta_{2}+\epsilon_{2})}:e^{-i\bar{\phi}(\gamma_{1})}e^{i\phi(\theta_{1})}:\right\rangle \right\rangle 
\end{eqnarray}
Again, the contraction in the last line contains one factor which
is independent of the integrated variables and reads:
\begin{eqnarray}
\mathcal{A}_{sasa} & = & \left\langle \left\langle e^{i\phi(\theta_{1}+i\pi+\epsilon_{1})}e^{i\phi(\theta_{2}+i\pi)}e^{i\phi(\theta_{2}+\epsilon_{2})}e^{i\phi(\theta_{1})}:\right\rangle \right\rangle \nonumber \\
 &  & G\left(\theta_{12}-\epsilon_{2}\right)G\left(\theta_{12}-i\pi\right)G\left(-i\pi-\epsilon_{1}\right)\label{eq:Asasa0}\\
 &  & G\left(-i\pi-\epsilon_{2}\right)G\left(\theta_{21}-i\pi-\epsilon_{1}\right)G\left(\theta_{21}-\epsilon_{1}\right)\nonumber 
\end{eqnarray}
on the other hand, the integral parts are as follows
\begin{eqnarray}
 &  & \intop_{\Gamma_{\sigma_{1}}^{(1)}}\frac{d\gamma_{1}}{2\pi}e^{A\gamma_{1}}W(\gamma_{1}+\frac{\theta}{2}-i\pi-\epsilon_{1})W(-\sigma_{1}(\frac{\theta}{2}+\gamma_{1}))W(\gamma_{1}-\frac{\theta}{2}-i\pi-\epsilon_{2})W(\gamma_{1}-\frac{\theta}{2})\\
 &  & \intop_{\mathcal{\Gamma}_{\sigma_{2}}^{(2)}}\frac{d\gamma_{2}}{2\pi}e^{A\gamma_{2}}W(\gamma_{2}+\frac{\theta}{2}-i\pi-\epsilon_{1})W(-\frac{\theta}{2}-\gamma_{2})W(\gamma_{2}-\frac{\theta}{2}-i\pi-\epsilon_{2})W(-\sigma_{2}(\gamma_{2}-\frac{\theta}{2}))\bar{G}(\gamma_{21})\nonumber 
\end{eqnarray}
with the contours depicted in figures \ref{fig:Contours-pmpm1} and
\ref{fig:Contours-pmpm2}.
The above formulas for the regularized diagonal form factor contain
a divergent integral whenever $k\ge1/2$. It is therefore necessary
to find a suitable and meaningful definition for these quantities.
We refer the interested reader to Appendix \ref{sec:Regularization},
where explicit formulas are provided.

\section{\label{sec:The-trace-of} The trace of the stress-energy tensor}

\subsection{The NLIE prediction}

As shown by Zamolodchikov \cite{Zamolodchikov:1989cf}, one can compute
the expectation value of the trace of the stress energy tensor for
any volume from the exact ground state energy. In order to compute
the latter, for the sine-Gordon theory, we can make use of the nonlinear
integral equation (NLIE) \cite{Klumper:1991b,Destri:1992qk,Destri:1994bv}:
\begin{eqnarray}
Z\left(\theta\right) & = & mR\sinh\theta-i\int_{-\infty}^{\infty}dxG_{0}(\theta-x-i\eta)\log\left(1+e^{iZ(x+i\eta)}\right)\nonumber \\
 &  & +i\int_{-\infty}^{\infty}dxG_{0}(\theta-x+i\eta)\log\left(1+e^{-iZ(x-i\eta)}\right)\label{eq:NLIE}
\end{eqnarray}
where $\eta<\pi\,\min(1,\xi)$
\begin{eqnarray}
G_{0}(\theta) & = & \int_{-\infty}^{\infty}\frac{dk}{2\pi}e^{ik\theta}\frac{\lyxmathsym{}\sinh\left(\frac{\pi(\xi-1)}{2}k\right)}{\sinh\frac{\pi\xi}{2}k\cosh\frac{\pi}{2}k}
\end{eqnarray}
and the ground state energy of the theory in finite volume $R$ can
be calculated using
\begin{equation}
E(R)=-m\:\Im\mathrm{m}\int_{-\infty}^{\infty}\frac{d\theta}{2\pi}\sinh(\theta+i\eta)\log\left(1+e^{iZ(\theta+i\eta)}\right)\label{eq:NLIE_energy}
\end{equation}
It satisfies
\begin{equation}
E(R)\rightarrow0\qquad\mathrm{as}\qquad R\rightarrow\infty\label{eq:NLIE_asympt}
\end{equation}
In the repulsive regime, one can continue analytically to $\eta=\pi$.
Introducing the functions \cite{Destri:1993qh} 
\begin{eqnarray}
\epsilon(\theta) & = & -iZ(\theta+i\pi)\nonumber \\
\bar{\epsilon}(\theta) & = & iZ(\theta-i\pi)
\end{eqnarray}
the ground state energy can be written as:
\begin{equation}
E(R)=-m\int\frac{d\theta}{2\pi}\cosh\theta\log(1+e^{-\varepsilon(\theta)})-m\int\frac{d\theta}{2\pi}\cosh\theta\log(1+e^{-\bar{\varepsilon}(\theta)})\label{eq:NLIEenergy}
\end{equation}
 The functions $\epsilon$, $\bar{\epsilon}$ are analogous to the
pseudoenergies of the thermodynamic Bethe ansatz approach and the
two terms can be thought of as resulting from the soliton-antisoliton
doublet. However, in contrast to TBA pseudoenergies, they are complex
valued functions. From (\ref{eq:NLIE}) it can be deduced that they
satisfy the equations

\begin{eqnarray}
\varepsilon(\theta) & = & mR\cosh\theta-\int dxG_{0}(\theta-x)\log(1+e^{-\varepsilon(x)})+\int dxG_{1}(\theta-x)\log(1+e^{-\bar{\varepsilon}(x)})\nonumber \\
\bar{\varepsilon}(\theta) & = & mR\cosh\theta-\int dxG_{0}(\theta-x)\log(1+e^{-\bar{\varepsilon}(x)})+\int dx\bar{G}_{1}(\theta-x)\log(1+e^{-\varepsilon(x)})\label{eq:NLIE_epsiloneqns}
\end{eqnarray}
where

\begin{eqnarray}
G_{1}(\theta) & = & G_{0}(\theta-i\pi)
\end{eqnarray}
Note that the function $G_{1}$ has a pole at $\theta=0$; on the
other hand, in the calculation of all physical quantities, this is
of no consequence. The two quantities $\varepsilon$ and $\bar{\varepsilon}$
are related by complex conjugation, as well as their derivatives below.
In fact, one can derive with respect to the volume and obtain
\begin{eqnarray}
\frac{1}{m}\frac{\partial\varepsilon}{\partial R}(\theta) & = & \cosh\theta+\int dxG_{0}(\theta-x)f(\theta)\frac{1}{m}\frac{\partial\varepsilon}{\partial R}(x)-\int dxG_{1}(\theta-x)\bar{f}(\theta)\frac{1}{m}\frac{\partial\bar{\varepsilon}}{\partial R}(x)
\end{eqnarray}
where the complex quantities $f(\theta)=\frac{1}{1+e^{\varepsilon(\theta)}}$
and $\bar{f}(\theta)=\frac{1}{1+e^{\bar{\varepsilon}(\theta)}}$ have
been used. On the other hand, the derivative of the first of (\ref{eq:NLIE_epsiloneqns})
with respect to the rapidity, denoted by a prime, is obtained as follows
\begin{equation}
\frac{1}{mR}\varepsilon'(\theta)=\sinh\theta+\int dxG_{0}(\theta-x)f(\theta)\frac{1}{mR}\varepsilon'(x)+\int dxG_{1}(\theta-x)\bar{f}(\theta)\frac{1}{mR}\bar{\varepsilon}'(x)
\end{equation}
It is then a simple matter to differentiate (\ref{eq:NLIEenergy})
and obtain

\begin{eqnarray}
\frac{\langle\Theta\rangle_{R}}{\langle\Theta\rangle_{\infty}}-1 & = & \frac{1}{m^{2}R}\frac{d}{dR}RE(R)\label{eq:thetaNLIE}\\
 & = & \int\frac{d\theta}{2\pi}\left\{ \cosh\theta\frac{1}{m}\left(\frac{\partial\varepsilon}{\partial R}(\theta)+\frac{\partial\bar{\varepsilon}}{\partial R}(\theta)\right)+\sinh\theta\frac{1}{mR}\left(\varepsilon'(\theta)+\bar{\varepsilon}'(\theta)\right)\right\} \nonumber \\
 & = & \frac{1}{2\pi}\Big(\intop d\theta(f+\bar{f})(\theta)+2\intop d\theta_{1}\intop d\theta_{2}\Big[f(\theta_{1})f(\theta_{2})G_{0}(\theta_{12})-f(\theta_{1})\bar{f}(\theta_{2})G_{1}(\theta_{12})\Big]\cosh\theta_{12}\nonumber \\
 &  & +2\intop d\theta_{1}\intop d\theta_{2}\intop d\theta_{3}\Big[f(\theta_{1})f(\theta_{2})f(\theta_{3})G_{0}(\theta_{12})-f(\theta_{1})f(\theta_{2})\bar{f}(\theta_{3})G_{0}(\theta_{12})G_{1}(\theta_{23})\nonumber \\
 &  & -f(\theta_{1})\bar{f}(\theta_{2})\bar{f}(\theta_{3})G_{1}(\theta_{12})G_{0}(\theta_{23})+f(\theta_{1})\bar{f}(\theta_{2})f(\theta_{3})G_{1}(\theta_{12})\bar{G}_{1}(\theta_{23})\Big]\cosh\theta_{13}+\ldots\Big)\nonumber 
\end{eqnarray}
The subtraction of $-1$ is related to the asymptotic property (\ref{eq:NLIE_asympt}),
which means that the ground state energy computed from the NLIE has
the bulk term subtracted. For the vacuum expectation value derived
from it, this entails that we obtain the finite size corrections to
the infinite volume value (\ref{eq:ThetaVEV},\ref{eq:vev-1}). For
later convenience, we also normalized the expectation value by its
infinite volume limit.

\subsection{Connected form factors for the trace of the stress-energy tensor\label{sub:Comparison}}

\begin{figure}
\centering{}\includegraphics[width=0.65\textwidth]{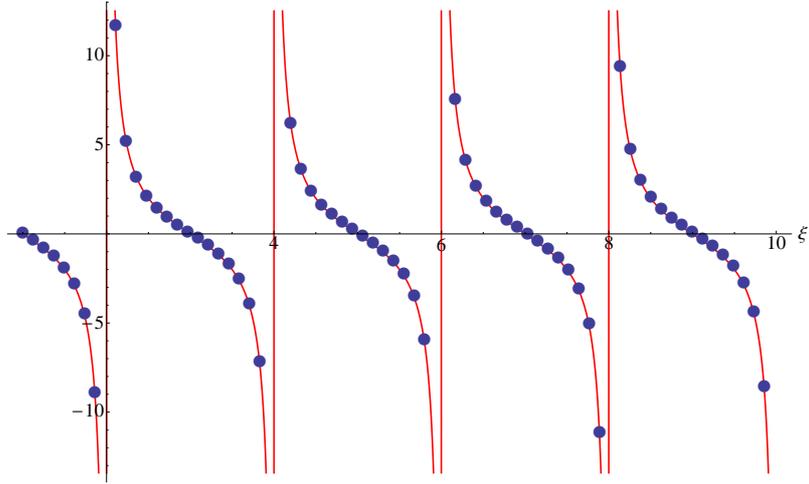}\caption{One-particle diagonal matrix element $\mathcal{F}_{s}^{\Theta}=\mathcal{F}_{a}^{\Theta}$
of the trace of the stress-energy tensor: comparison between the regularization
procedure outlined in Section \ref{sub:The-diagonal-form} (dots)
against the analytic formula obtained from Lukyanov's form factor
(solid line). The normalization through the operator vacuum expectation
value is not included in this plot.}
\label{fig:Compare1p}
\end{figure}

Diagonal matrix elements of the trace of the stress-energy tensor
can be computed as outlined in \cite{Leclair:1999ys}, which is reviewed
in Appendix \ref{sec:Regularization}. In particular, the diagonal
one-particle form factor is given by 
\begin{equation}
\mathcal{F}_{s}^{\Theta}=\mathcal{F}_{a}^{\Theta}=2\cot\frac{\pi\xi}{2}\mathcal{G}_{1}\label{eq:2pconnff}
\end{equation}
Comparing this to the result (\ref{eq:Fsa_connected}) gives a first
check of the regularization method, which is shown in Figure \ref{fig:Compare1p}.

\begin{figure}[H]
\includegraphics[width=0.5\textwidth]{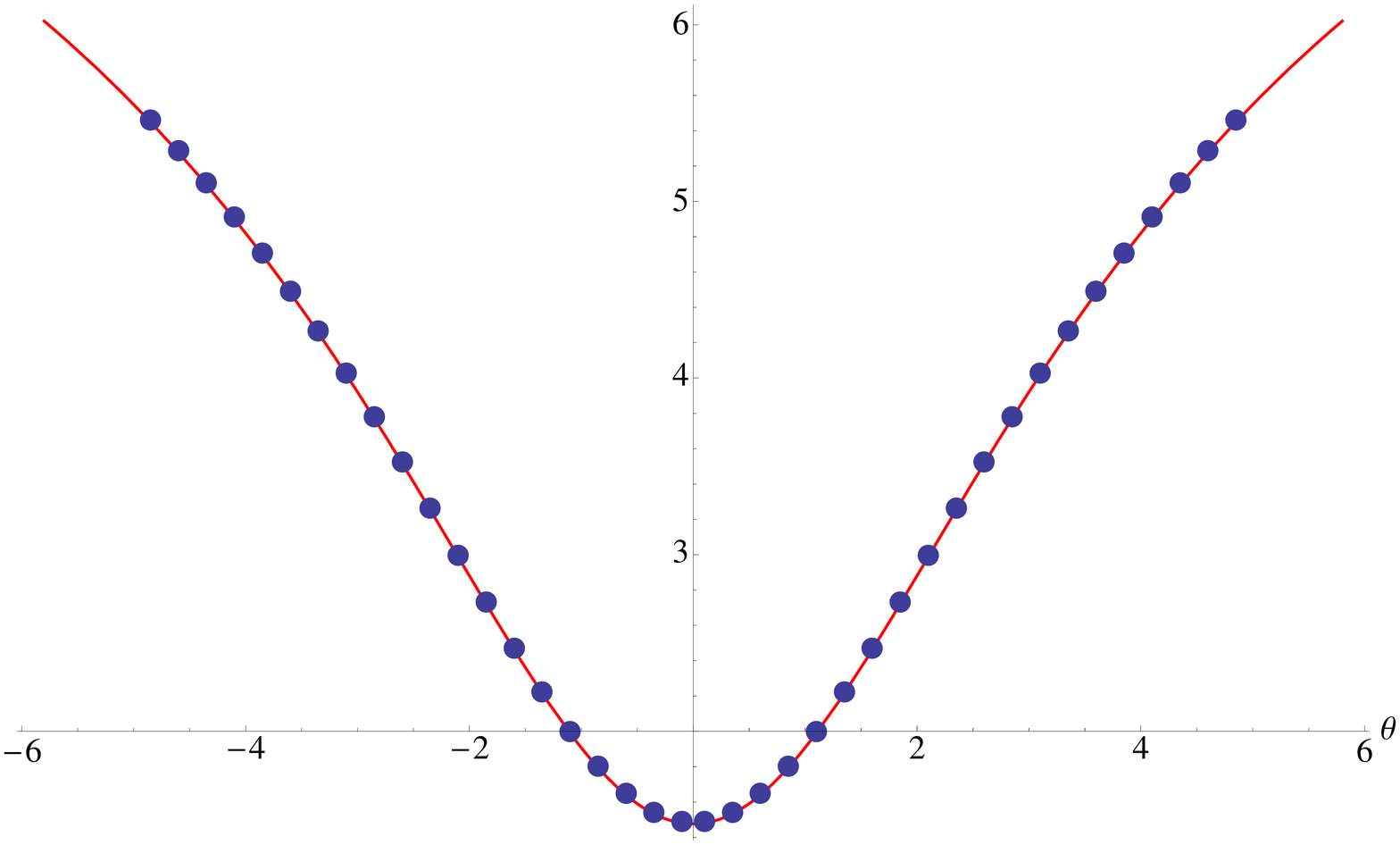}\includegraphics[width=0.5\textwidth]{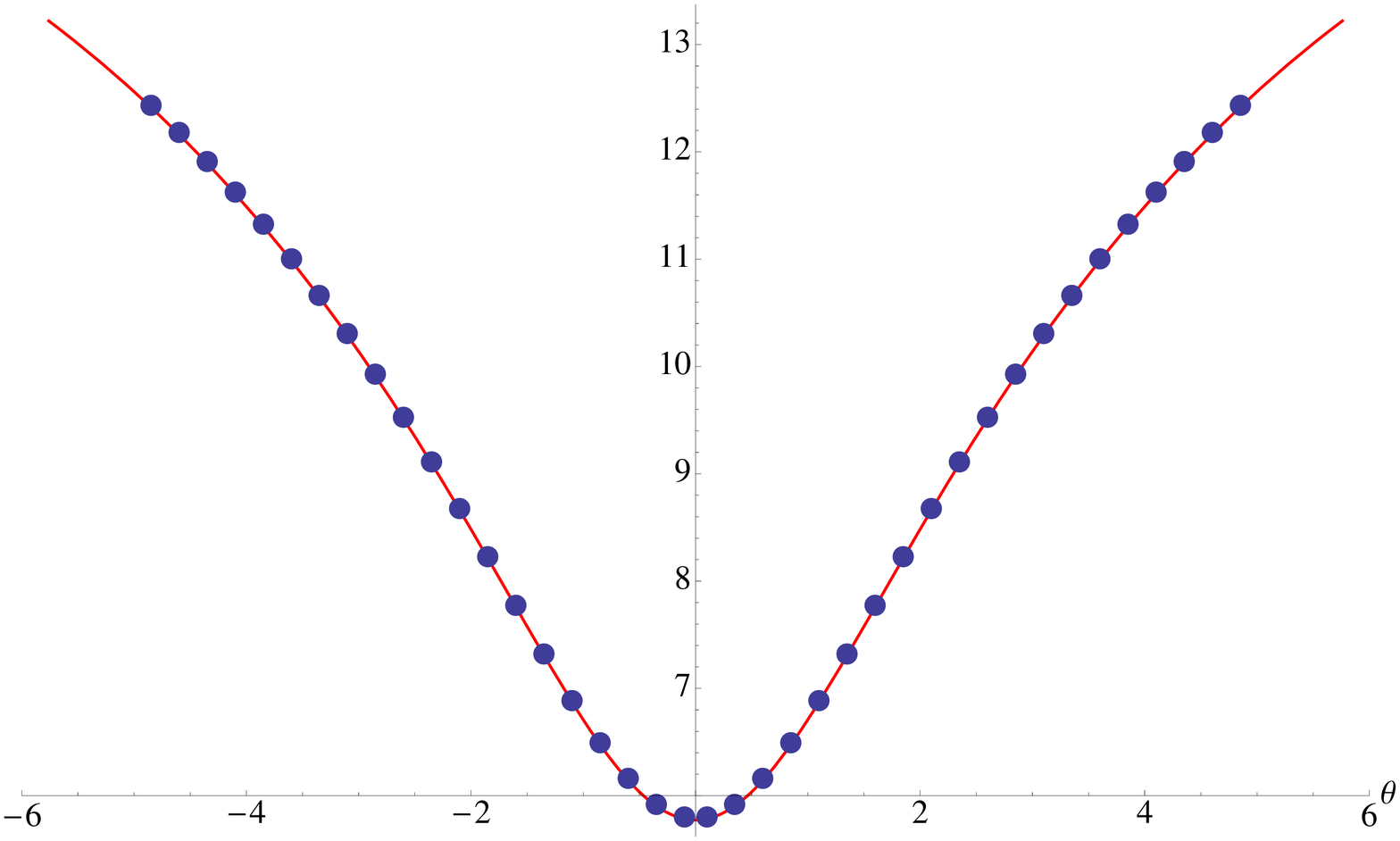}\caption{\label{fig:Left:-connected-diagonal}Left: connected diagonal matrix
element $\mathcal{F}_{ss}^{\Theta}=\mathcal{F}_{aa}^{\Theta}$. Right:
connected diagonal matrix element $\mathcal{F}_{sa}^{\Theta}+\mathcal{F}_{as}^{\Theta}$
for $\xi=1.6129$. The solid line corresponds to the formulas (\ref{eq:connFFssaa_sasa}),
while the dots are evaluated using Appendix \ref{sec:Regularization}. }
\end{figure}

Moreover, the connected diagonal two-particle form factors are also
explicitly known (see Appendix \ref{sec:The-connected-diagonal}):

\begin{eqnarray}
 &  & \mathcal{F}_{ss}^{\Theta(c)}(\theta_{1},\theta_{2})=-8\pi\mathcal{G}_{1}\cot\frac{\pi\xi}{2}G_{0}(\theta_{12})\cosh\theta_{12}\label{eq:connFFssaa_sasa}\\
 &  & \mathcal{F}_{aa}^{\Theta(c)}(\theta_{1},\theta_{2})=-8\pi\mathcal{G}_{1}\cot\frac{\pi\xi}{2}G_{0}(\theta_{12})\cosh\theta_{12}\nonumber \\
 &  & (\mathcal{F}_{sa}^{\Theta}+\mathcal{F}_{as}^{\Theta})^{(c)}(\theta_{1},\theta_{2})=4\pi\mathcal{G}_{1}\cot\frac{\pi\xi}{2}(G_{1}+\bar{G}_{1}-2G_{0})(\theta_{12})\cosh\theta_{12}\nonumber 
\end{eqnarray}
We checked (for values $1/2<\xi<2$, where we need to take into account
only principal poles) that the connected form factors as obtained
in the form (\ref{eq:Fsasaconn},\ref{eq:Fssaaconn}) from Lukyanov's
expressions (as given in Appendix \ref{sec:Regularization}) agree
with these (see figure \ref{fig:Left:-connected-diagonal}). In addition
to that, the two can be compared with the results from the regularization
procedure of \cite{Palmai:2011kb}, again resulting in agreement among
the results, thus providing a threefold check.

For future reference, we also write here the soliton-breather and
the two-breather connected diagonal matrix element of the trace of
the stress-energy tensor:
\begin{eqnarray}
\mathcal{F}_{b\mp}^{\Theta(c)}(\theta_{1},\theta_{2}) & = & -16\pi\mathcal{G}_{1}\sin\frac{\pi\xi b}{2}\cot\frac{\pi\xi}{2}G_{sb}(\theta_{12})\cosh\theta_{12}\label{eq:connFFsbba}\\
\mathcal{F}_{b_{1}b_{2}}^{\Theta(c)}(\theta_{1},\theta_{2}) & = & -32\pi\mathcal{G}_{1}\sin\frac{\pi\xi b_{1}}{2}\sin\frac{\pi\xi b_{2}}{2}\cot\frac{\pi\xi}{2}G_{b_{1}b_{2}}(\theta_{12})\cosh\theta_{12}\label{eq:connFFbbbb}
\end{eqnarray}

\subsection{Comparing the NLIE to the series}

Let us start with an analytical comparison in the repulsive regime
$\xi>1$. It is easy to see that expanding the exact result (\ref{eq:thetaNLIE})to
second order in $e^{-mR}$ of produces exactly the terms of series
(\ref{eq:expansion}) up to this order.

\begin{figure}[H]
\includegraphics[width=0.5\textwidth]{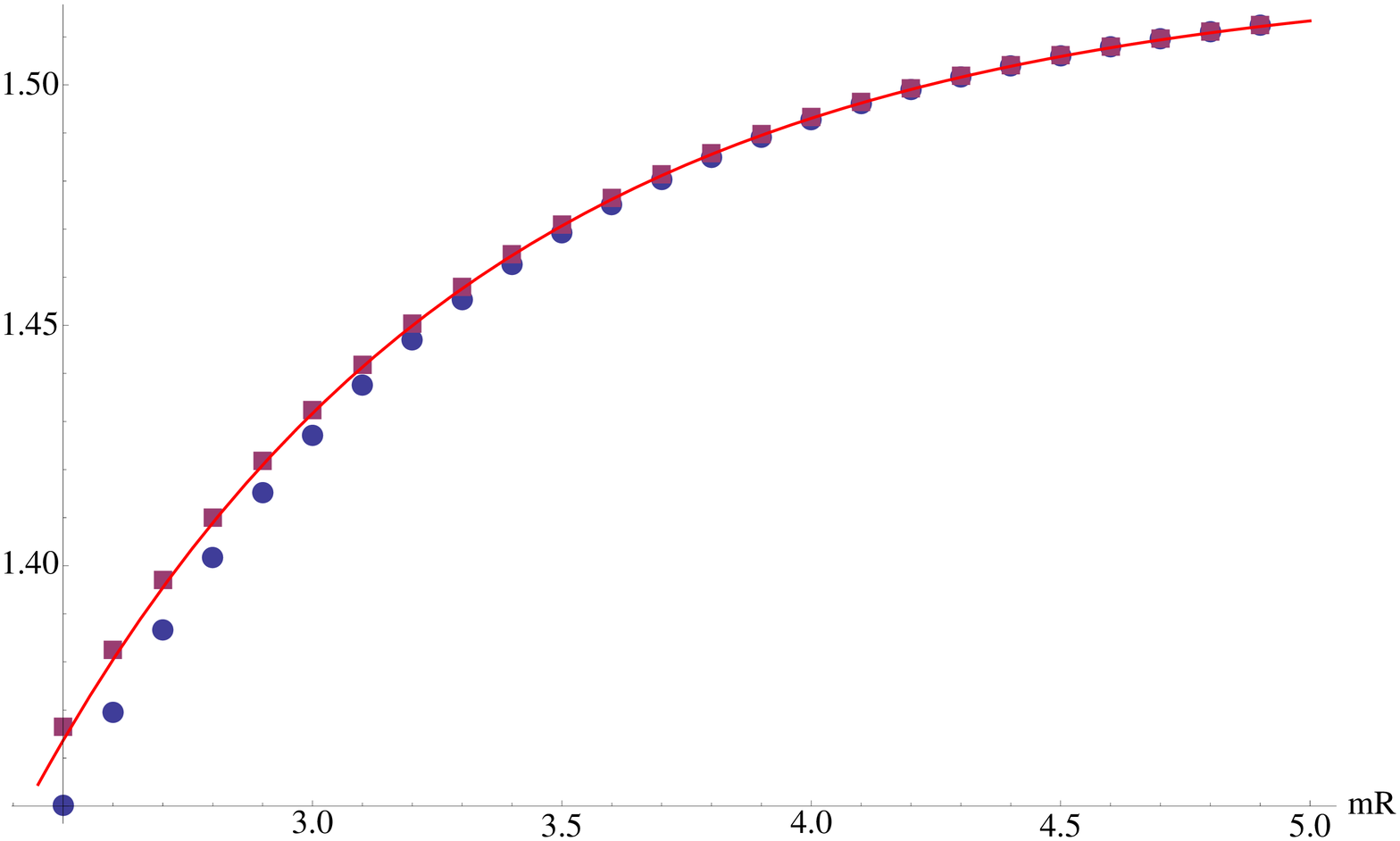}\includegraphics[width=0.5\textwidth]{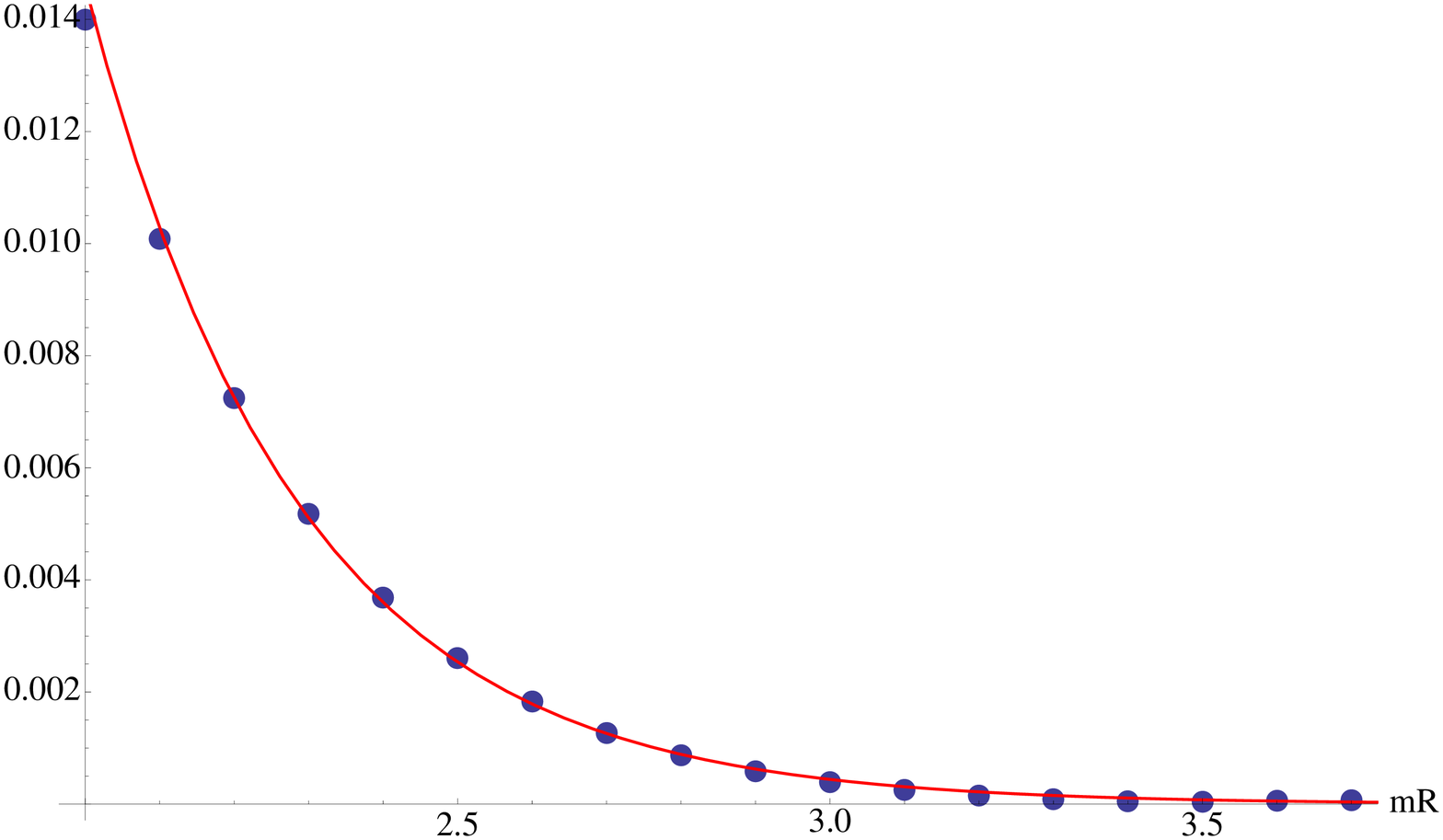}

\caption{Left: Form factor expansion to first (blue circles) and second (purple
squares) order in $e^{-mR}$ of the trace of the stress-energy tensor
at $\xi=1.6129$, as a function of the inverse temperature $R$, compared
to the exact value (red solid line). Right: exponential decay of the
deviation of the expansion from the exact value, the exponent being
$\sim3.4mR$.\label{fig:vevThetaR090}}
\end{figure}

First we need to expand the expression (\ref{eq:NLIE_epsiloneqns})
and the relative (complex) filling factor, which leads to:
\begin{eqnarray}
\varepsilon(\theta) & \simeq & mR\cosh\theta-\int dx\left(G_{0}(\theta-x)-G_{1}(\theta-x)\right)e^{-mR\cosh x}+O(e^{-2mR})\nonumber \\
f\left(\theta\right) & \simeq & e^{-mR\cosh\theta}+\int dx\left(G_{0}(\theta-x)-G_{1}(\theta-x)\right)e^{-mR\left(\cosh\theta+\cosh x\right)}-e^{-2mR\cosh\theta}\nonumber \\
 &  & +O(e^{-3mR})
\end{eqnarray}
while the one for $\bar{\varepsilon}$ and $\bar{f}$ are obtained
by the substitution $G_{1}\to\bar{G}_{1}$. The expansion of the NLIE
result gives:

\begin{eqnarray}
\text{\ensuremath{\frac{\langle\Theta\rangle_{R}}{\langle\Theta\rangle_{\infty}}}}-1 & = & \frac{1}{2\pi}\Big(\sum_{j=s,a}\intop d\theta e^{-mR\text{\ensuremath{\cosh}}\theta}-\sum_{j=s,a}\intop d\theta e^{-2mR\text{\ensuremath{\cosh}}\theta}\nonumber \\
 &  & +\intop d\theta_{1}\intop d\theta_{2}e^{-mR\text{\ensuremath{\cosh}}\theta_{1}-mR\text{\ensuremath{\cosh}}\theta_{2}}\left(2G_{0}(\theta_{21})-\left(G_{1}(\theta_{21})+\bar{G}_{1}(\theta_{21})\right)\right)\nonumber \\
 &  & +\intop d\theta_{1}\intop d\theta_{2}e^{-mR\text{\ensuremath{\cosh}}\theta_{1}-mR\text{\ensuremath{\cosh}}\theta_{2}}\left(2G_{0}(\theta_{21})-G_{1}(\theta_{21})-\bar{G}_{1}(\theta_{21})\right)\cosh\theta_{21}\nonumber \\
 &  & +O(e^{-3mR})\Big)
\end{eqnarray}
Using expressions (\ref{eq:2pconnff},\ref{eq:connFFssaa_sasa}) for
the connected form factors of the trace of the energy-momentum tensor,
one can easily recognize that this is indeed identical to the solitonic
terms in (\ref{eq:expansion}), once the normalization of the two-particle
form factor according to (\ref{eq:2pconnff}) is taken into account.
Note that this also validates, using this operator as an example,
the conjecture stated in \cite{Palmai:2012kf}, for which only heuristic
argument and numerical support was obtained in the original paper.
Convergence of the series is illustrated in Figure \ref{fig:vevThetaR090},
in which we compare the expansion for the trace of the stress-energy
tensor to the NLIE data obtained by a recursive solution of (\ref{eq:NLIE}).

\begin{figure}[H]
\begin{centering}
\includegraphics[width=0.5\textwidth]{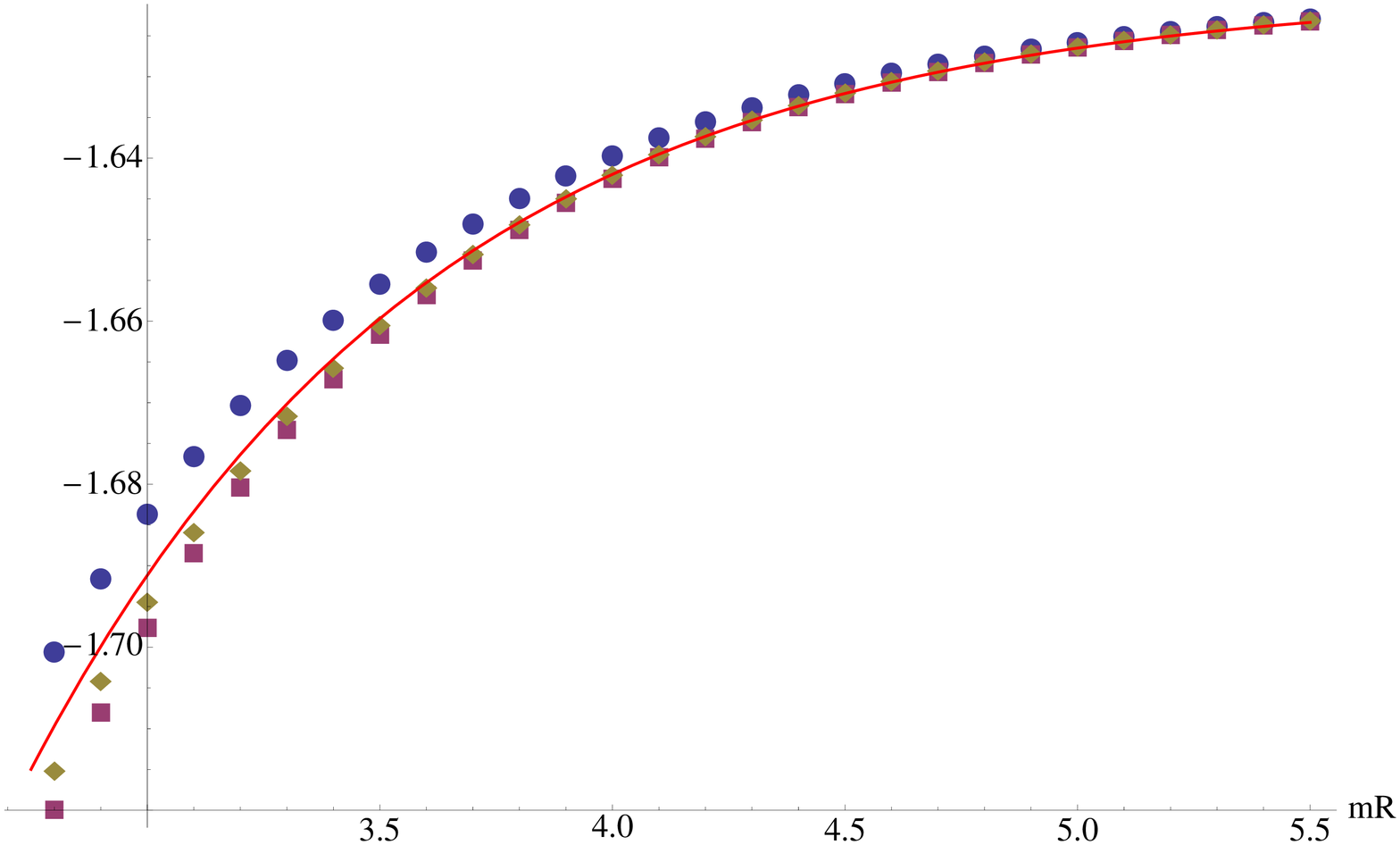}\includegraphics[width=0.5\textwidth]{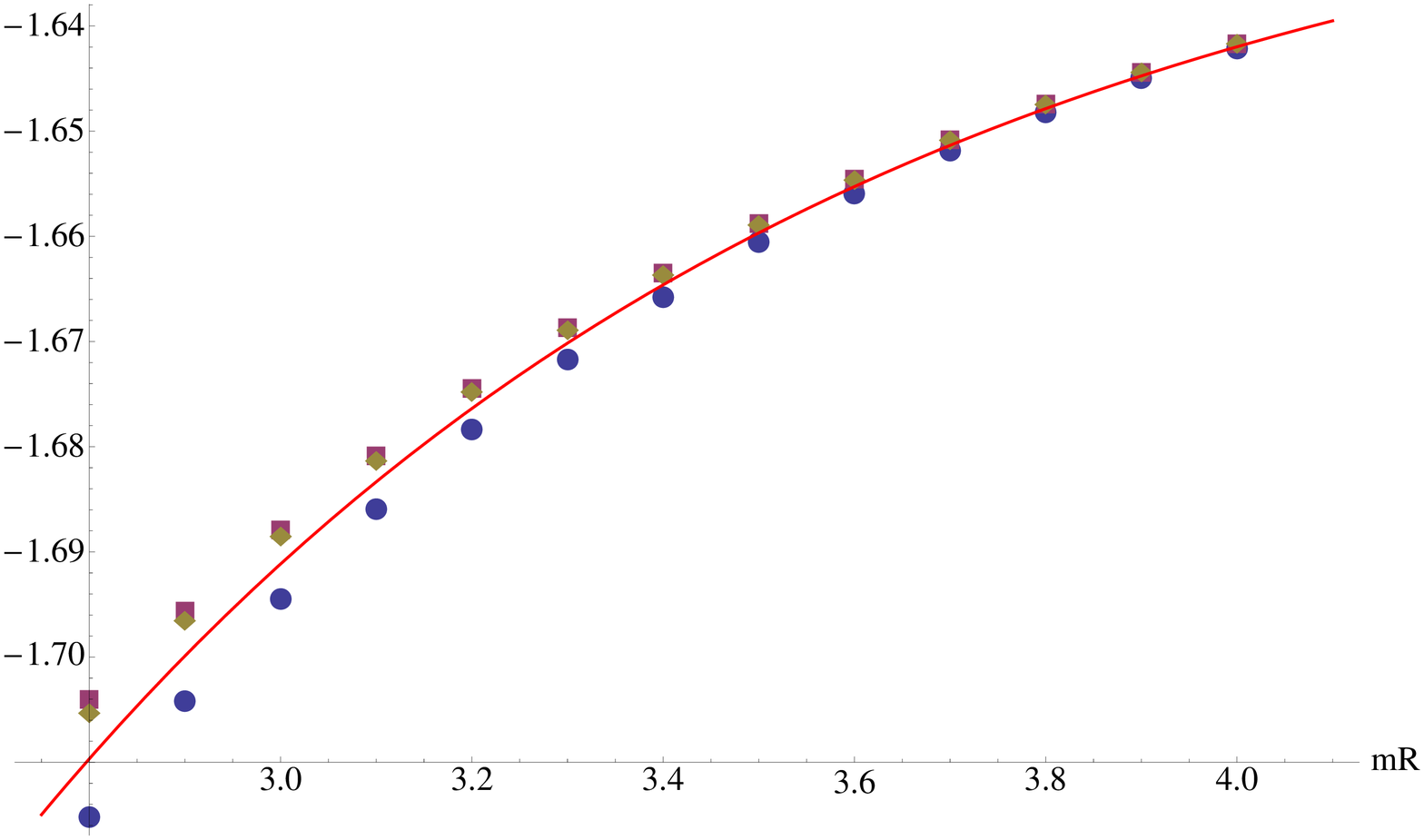}\\
\includegraphics[width=0.5\textwidth]{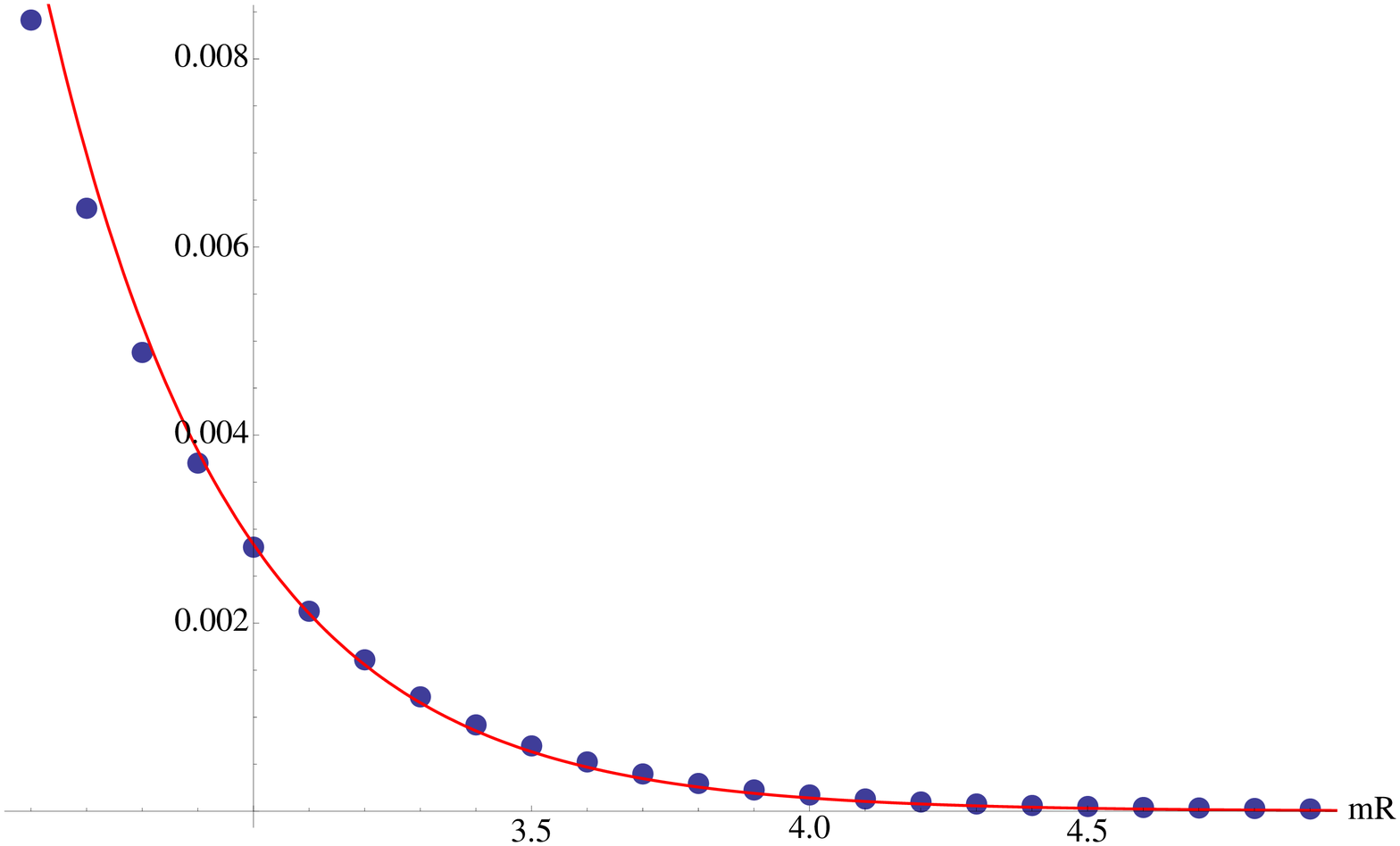}
\par\end{centering}

\caption{Left: Form factor expansion up to one soliton (blue circles), up to
one breather (purple squares) and up to two solitons (yellow rhombuses)
of the trace of the stress-energy tensor at $\xi=0.531915$, as a
function of the inverse temperature $R$, compared to the exact value
(red solid line). Right: inclusion of the contributions up to two
solitons (blue circles), one soliton and one breather (purple squares)
and up to two breathers (yellow rhombuses). Below: exponential decay
of the deviation of the expansion from the exact value, compared to
$e^{-3mR}$ decay.\label{fig:vevThetaR120}}
\end{figure}

The analytic continuation of the counting function which underlies
formula (\ref{eq:thetaNLIE}) to imaginary values of its argument
has to take into account that the first determination strip of the
$G_{1}$ function has a width of $\pi\,\min(1,\xi)$. This implies
that the so-called second determination must be used in the attractive
regime; moreover the series must also be partially summed to obtain
the expected exponential decay with exponent determined from the breather
mass. To avoid these complications, we resort to numerical calculations,
and in figure \ref{fig:vevThetaR120} it is shown that that the exact
expectation value of $\Theta$ is well reproduced by the conjectured
expansion. We can also provide the finite temperature corrections
to the vacuum expectation value of all the operator $\left\langle e^{ik\beta\phi}\right\rangle _{R}$
for integer $k$. Their soliton-antisoliton form factors are finite
in the diagonal limit and can be straightforwardly derived from \cite{Lukyanov:1997bp}
\begin{equation}
\mathcal{F}_{s/a}^{k}=(-1)^{k+1}\mathcal{G}_{k}\sum_{m=1}^{k}\left(4m-2\right)\prod_{l=m-k,k\ne0}^{m+k-1}\cot\frac{\pi\xi k}{2}
\end{equation}
for $k=1,2\dots$, while our computation provides the two-particle
connected diagonal matrix elements in (\ref{eq:Fsasaconn},\ref{eq:Fssaaconn}).

However, at present we can not make a useful comparison to an independent
calculation. The NLIE only provides access to the cases $k=\pm1$,
i.e. the trace of the stress energy tensor. For higher $k$, one could
resort to numerical determination using truncated conformal space
approach (TCSA), originally invented by Yurov and Zamolodchikov \cite{Yurov:1989yu}
and extended to perturbations of the free boson theory in \cite{Feverati:1998va}.
However, the evaluation of matrix elements is plagued by ultraviolet
differences. Using the results in \cite{Szecsenyi:2013gna}, it can
be easily seen that the matrix elements of $e^{ik\beta\phi}$ are
only finite in TCSA for $\xi<1/(2|k|-1)$. Even for $k=\pm2$ this
is deep in the attractive regime, with numerous light breather states.
For these couplings, before getting to the interesting novel part
of our expansion which involves non-diagonal scattering (i.e. the
two-soliton states), a lot of corrections coming from states with
diagonal scattering must be summed over. Besides this being a very
tedious task, the part we would really wish to put to the test would
be so tiny as to escape any useful comparison.

An alternative possibility is to renormalize TCSA to obtain the expectation
values at a less attractive, or even repulsive coupling. Such renormalization
has already been performed for energy levels \cite{Giokas:2011ix,Beria:2013hz}.
By extending the methods of \cite{Szecsenyi:2013gna} it should be
possible to extend the procedure to expectation values. However, a
preliminary investigation that we performed indicates that this is
a very nontrivial task, and is clearly out of the scope of the present
work. We hope to return to this issue in the future.

\section{Conclusions}

In this work we have studied the low-temperature expansion for one-point
functions in sine-Gordon model, which can be considered as a paradigmatic
example of integrable field theory with non-diagonal scattering. Following
the ideas of Pozsgay and Takács \cite{Pozsgay:2007gx}, we proposed
a series expansion which enables to compute the vacuum expectation
value of exponential operators. The formula is expressed in terms
of connected components of the diagonal matrix elements of the operator
and can be considered as a generalization of the LeClair-Mussardo
expansion \cite{Leclair:1999ys} to non-diagonal scattering. However,
in contrast with the latter and with the established formalism in
diagonal theories, we could not write our result in terms of single
particle energies and momenta dressed by the thermodynamic Bethe ansatz.
In this respect, there remains a marked difference between diagonal
and non-diagonal scattering theories. Our results were verified for
the case of the trace of the stress-energy tensor by comparing against
the NLIE approach.

We have also developed a way to evaluate the connected diagonal matrix
elements analytically, starting from the integral expressions of the
form factors obtained by Lukyanov \cite{Lukyanov:1997bp}. For a special
value of the exponent, the form factors were compared to those of
the trace of the stress-energy tensor which can be obtained by different
routes, providing a nontrivial validation of the procedure.

Finally, we have provided analytic support to the relation between
finite and infinite-volume form factors conjectured in \cite{Palmai:2012kf}
which had previously been supported only by intuitive reasoning and
numerical evidence.

An interesting open problem is to extend our results to states containing
more than two solitons/antisolitons, which would then enable the explicit
evaluation of the higher terms in the proposed series expansion.

\subsection*{Acknowledgments}

This work was partially supported by OTKA 81461 and Lendület LP2012-50/2012
grants. F.B. acknowledges support from Ministry of Science, Technology
and Innovation of Brazil.

\appendix
\makeatletter \renewcommand{\theequation}{\hbox{\normalsize\Alph{section}.\arabic{equation}}} 
\@addtoreset{equation}{section} 
\renewcommand{\thefigure}{\hbox{\normalsize\Alph{section}.\arabic{figure}}} 
\@addtoreset{figure}{section} 
\renewcommand{\thetable}{\hbox{\normalsize\Alph{section}.\arabic{table}}} 
\@addtoreset{table}{section} \makeatother

\section{The diagonal matrix elements of the trace of the stress-energy tensor\label{sec:The-connected-diagonal}}

Here we summarize the calculation, introduced in \cite{Leclair:1999ys,Mussardo},
of the diagonal matrix elements of the operator $\Theta=T_{\mu}^{\mu}$.
First we recall that for a given local operator $\mathcal{O}$, the
form factor dependence under space and time translations can be written
as
\begin{equation}
F^{\mathcal{O}(x,t)}(\theta_{1},\ldots,\theta_{n})=e^{-ix\left(\sum_{j}m_{j}\sinh\theta_{j}\right)+it\left(\sum_{j}m_{k}\cosh\theta_{j}\right)}F^{\mathcal{O}(0,0)}(\theta_{1},\ldots,\theta_{n})\label{eq:translation}
\end{equation}
in which the energy and momentum of the $j$-th particle, having mass
$m_{j}$, have been parametrized as $e_{j}=m_{j}\cosh\theta_{j}$
and $p_{j}=m_{j}\sinh\theta_{j}$, respectively.

Conservation of the stress tensor implies that: 
\begin{equation}
T_{\mu\nu}=\left(\partial_{\mu}\partial_{\nu}-g_{\mu\nu}\square\right)A
\end{equation}
for some scalar field $A$. Knowing the form factors of this field,
as well as the property (\ref{eq:translation}), allows to compute
those of $\Theta$ by the use of:
\begin{eqnarray}
\mathcal{F}^{\Theta}\left(\theta_{1},\ldots,\theta_{n}\right) & = & \lim_{\epsilon_{1}\ldots\epsilon_{n}\to0}F^{(\partial_{1}^{2}-\partial_{0}^{2})A}(\theta_{1}+i\pi+\epsilon_{1},\ldots\theta_{n}+i\pi+\epsilon_{n},\theta_{n},\ldots,\theta_{1})\label{eq:AtoTheta}\\
 & = & -\lim_{\epsilon_{1}\ldots\epsilon_{n}\to0}\sum_{j,k}\epsilon_{j}\epsilon_{k}m_{j}m_{k}\cosh\left(\theta_{j}-\theta_{k}\right)F^{A}(\theta_{1}+i\pi+\epsilon_{1},\ldots,\theta_{n}+i\pi+\epsilon_{n},\theta_{n},\ldots,\theta_{1})\nonumber 
\end{eqnarray}
where the overall normalization $\mathcal{N}$ is left undetermined. 

Following the procedure explained in \cite{Mussardo}, one can determine
the two-particle form factor from the expectation value of the Hamiltonian
(the $T_{00}$ component of the stress energy tensor)

\begin{equation}
\left\langle \theta+\epsilon\left|\intop\frac{dx_{1}}{2\pi}T_{00}\left(x_{1}\right)\right|\theta\right\rangle =-m\frac{\left(\sinh\theta-\sinh\left(\theta+\epsilon\right)\right)^{2}}{\cosh\theta}\delta\left(\epsilon\right)\mathcal{F}^{A}
\end{equation}
by comparing it with the single particle energy 
\begin{equation}
\left\langle \theta+\epsilon\left|\intop\frac{dx_{1}}{2\pi}T_{00}\left(x_{1}\right)\right|\theta\right\rangle =2\pi\delta\left(\epsilon\right)m\cosh\theta
\end{equation}
The above formula (\ref{eq:translation}) implies that the behavior
of the two particle form factor is
\begin{equation}
F_{as}^{A}(\theta+i\pi+\epsilon,\theta)=F_{sa}^{A}(\theta+i\pi+\epsilon,\theta)\simeq-\frac{2\pi}{\epsilon^{2}}\label{eq:FAas}
\end{equation}
in the diagonal limit $\epsilon\to0$. In order to match Lukyanov's
normalization \cite{Lukyanov:1997bp} of the exponential operator,
an overall normalization has to be left undetermined. This normalization
can be determined by comparison with the exact formula (\ref{eq:2pconnff}).
Analogous reasoning and the expression (\ref{eq:BreatherMass}) allows
one to compute the breather diagonal matrix elements.

Once the proper normalization factor is fixed, higher form factors
are uniquely determined and can be computed recursively the by repeated
use of the kinematical pole equation, which encodes the singular part
of the function when $\theta_{m}\to\theta_{n}+i\pi$:
\begin{eqnarray}
iF_{j_{1}\ldots j_{n}}\left(\theta_{1},\ldots\theta_{m},\ldots,\theta_{n}\right) & \simeq & C_{j_{n}k_{n}}\frac{1}{\theta_{m}-\theta_{n}-i\pi}F_{k_{1}\ldots\hat{k}_{m}\ldots k_{n-1}}\left(\theta_{1},\ldots\hat{\theta}_{m}\ldots,\theta_{n-1}\right)\\
 &  & \Big[\delta_{j_{1}}^{k_{1}}\ldots\delta_{j_{1}}^{k_{1}}S_{c_{1}j_{n-1}}^{k_{m}k_{n-1}}\left(\theta_{m}-\theta_{n-1}\right)\ldots S_{j_{m}j_{m+1}}^{c_{n-m-2}k_{m+1}}\left(\theta_{m}-\theta_{m+1}\right)\nonumber \\
 &  & -e^{2\pi i\omega_{\mathcal{O}\Psi}}S_{j_{1}c_{1}}^{k_{1}k_{m}}\left(\theta_{1}-\theta_{m}\right)\ldots S_{j_{m-1}j_{m}}^{k_{m-1}c_{m-2}}\left(\theta_{m-1}-\theta_{m}\right)\delta_{j_{m+1}}^{k_{m+1}}\ldots\delta_{j_{n-1}}^{k_{n-1}}\Big]\nonumber 
\end{eqnarray}
where $C$ is the charge conjugation matrix, which in the case of
sine-Gordon is the Pauli matrix $\sigma^{x}$ in the soliton-antisoliton
sector, while it is the identity in the breather sector. The mutual
locality factor $\omega_{\mathcal{O}\Psi}$ encodes the braiding properties
of the operator $\mathcal{O}$ with the field $\Psi$ that interpolates
particles \cite{Lukyanov:1993pn}. One needs to select all the contributions
which diverge as $O\left(\frac{1}{\epsilon_{j}\epsilon_{k}}\right)$,
which will give a finite contribution when inserted into (\ref{eq:AtoTheta}).

In particular, this procedure applied to (\ref{eq:FAas}) yields the
results (\ref{eq:connFFssaa_sasa})
\begin{eqnarray}
 &  & \mathcal{F}_{ss}^{\Theta(c)}(\theta_{1},\theta_{2})=\mathcal{F}_{aa}^{\Theta}(\theta_{1},\theta_{2})=-8\pi\mathcal{G}_{1}\cot\frac{\pi\xi}{2}G_{0}(\theta_{12})\cosh\theta_{12}\label{eq:connFFssaa_sasa-1}\\
 &  & (\mathcal{F}_{sa}^{\Theta(c)}+\mathcal{F}_{as}^{\Theta})(\theta_{1},\theta_{2})=4\pi\mathcal{G}_{1}\cot\frac{\pi\xi}{2}(G_{1}+\bar{G}_{1}-2G_{0})(\theta_{12})\cosh\theta_{12}\nonumber 
\end{eqnarray}
in the solitonic sector using the S-matrix (\ref{eq:Smatrix}). On
the other hand, using (\ref{eq:FAas}), the breather mass (\ref{eq:BreatherMass})
and the (diagonal) breather-breather and soliton-breather S-matrices
\cite{zam-zam}, one obtains (\ref{eq:connFFsbba},\ref{eq:connFFbbbb}):
\begin{eqnarray}
\mathcal{F}_{b\mp}^{\Theta(c)}(\theta_{1},\theta_{2}) & = & -16\pi\mathcal{G}_{1}\sin\frac{\pi\xi b}{2}\cot\frac{\pi\xi}{2}G_{sb}(\theta_{12})\cosh\theta_{12}\label{eq:connFFsbba-1}\\
\mathcal{F}_{b_{1}b_{2}}^{\Theta(c)}(\theta_{1},\theta_{2}) & = & -32\pi\mathcal{G}_{1}\sin\frac{\pi\xi b_{1}}{2}\sin\frac{\pi\xi b_{2}}{2}\cot\frac{\pi\xi}{2}G_{b_{1}b_{2}}(\theta_{12})\cosh\theta_{12}\label{eq:connFFbbbb-1}
\end{eqnarray}

\section{The connected diagonal soliton form factors of the exponential field\label{sec:Regularization}}

We focus here on the four-particle diagonal matrix element. We need
to collect the parts which stay finite when $\epsilon_{1,2}\to0$.
According to the considerations in section \ref{sub:Form-factors-review},
all the form factors in the soliton sector share the same structure:
there is an overall factor, which depends only on the rapidities but
not on the particle species, and a double integral.

The double integral is generally time-consuming to evaluate numerically.
It has the following form: two functions $g$ and $h$, both of which
depend only on one of the two integration variables, multiplying $\bar{G}$,
which instead depends on the difference of the two. Therefore, the
double integral can be written as a sum of products of two simple
integrals by a simple trick \cite{Palmai:2011kb}: 
\begin{eqnarray}
I_{ss} & = & \intop d\gamma_{1}\intop d\gamma_{2}g\left(\gamma_{1}\right)h\left(\gamma_{2}\right)\bar{G}\left(\gamma_{1}-\gamma_{2}\right)\label{eq:deompIss}\\
 & = & -\frac{\mathcal{C}_{2}\xi}{16}\sum_{\alpha_{1},\alpha_{2}=\pm}\alpha_{1}\alpha_{2}e^{i\alpha_{2}\pi/\xi}\intop_{\Gamma^{(1)}}d\gamma_{1}e^{\left(\alpha_{1}+\alpha_{2}/\xi\right)\gamma_{1}}g\left(\gamma_{1}\right)\intop_{\Gamma^{(2)}}d\gamma_{2}h\left(\gamma_{2}\right)e^{-\left(\alpha_{1}+\alpha_{2}/\xi\right)\gamma_{2}}\nonumber \\
I_{sa} & = & \intop d\gamma_{1}\intop d\gamma_{2}g\left(\gamma_{1}\right)h\left(\gamma_{2}\right)\bar{G}\left(\gamma_{1}-\gamma_{2}-i\pi\right)\nonumber \\
 & = & -\frac{\mathcal{C}_{2}\xi}{16}\sum_{\alpha_{1},\alpha_{2}=\pm}\alpha_{1}\alpha_{2}\intop_{\Gamma^{(1)}}d\gamma_{1}e^{\left(\alpha_{1}+\alpha_{2}/\xi\right)\gamma_{1}}g\left(\gamma_{1}\right)\intop_{\Gamma^{(2)}}d\gamma_{2}e^{-\left(\alpha_{1}+\alpha_{2}/\xi\right)\gamma_{2}}h\left(\gamma_{2}\right)\label{eq:decompIsa}
\end{eqnarray}
using the definition (\ref{eq:G}).

Note that the integrals are evaluated along the contours $\Gamma^{(1,2)}$,
which will be deformed to either the real axis or the lines $\Im m\gamma_{1}=\pi$
(as for the one in figure \ref{fig:Contours-pmpm2}). In the course
of this deformation some poles are encountered; in the regime $1/2<\xi<2$,
one only needs to treat principal poles of the $W$ functions (\ref{eq:defW}).
For each integral, there is a contribution of order $\frac{1}{\epsilon_{1}}$,
which we write as $\frac{1}{\epsilon_{1}}P_{1}^{(j)}\left(\theta,\epsilon_{1},\epsilon_{2}\right)$,
and another one of order $\frac{1}{\epsilon_{2}}$, written as $\frac{1}{\epsilon_{2}}P_{2}^{(j)}\left(\theta,\epsilon_{1},\epsilon_{2}\right)$.
The index $j=1,2$ represents the integral from which the given contribution
originates. We postpone for the moment the analysis of the functions
$P$, by just remarking that they depend on the difference of rapidities
$\theta=\theta_{2}-\theta_{1}$ only.

From this, one has one family of form factors, in the form:
\begin{eqnarray}
 &  & F_{ssaa}\left(\theta_{1}+i\pi+\epsilon_{1},\theta_{2}+i\pi+\epsilon_{2},\theta_{2},\theta_{1}\right)=\nonumber \\
 &  & \mathcal{G}_{k}\mathcal{A}_{ssaa}\left(\theta,\epsilon_{1},\epsilon_{2}\right)\sum_{\sigma_{1}\sigma_{2}}\sigma_{1}e^{i\sigma_{1}\pi(1+1/\xi)/2}\sigma_{2}e^{i\sigma_{2}\pi(1+1/\xi)/2}\sum_{\alpha_{1}\alpha_{2}}\alpha_{1}\alpha_{2}\nonumber \\
 &  & e^{i\alpha_{2}\pi/\xi}\left(\frac{P_{1,\sigma_{1}\alpha_{1}\alpha_{2}}^{(1)}\left(\theta,\epsilon_{1},\epsilon_{2}\right)}{\epsilon_{1}}+\frac{P_{2,\sigma_{1}\alpha_{1}\alpha_{2}}^{(1)}\left(\theta,\epsilon_{1},\epsilon_{2}\right)}{\epsilon_{2}}+I_{\sigma_{1}\alpha_{1}\alpha_{2}}^{(1)}\left(\theta,\epsilon_{1},\epsilon_{2}\right)\right)\nonumber \\
 &  & \Big(\frac{P_{1,\sigma_{1}\alpha_{1}\alpha_{2}}^{(2)}\left(\theta,\epsilon_{1},\epsilon_{2}\right)}{\epsilon_{1}}+\frac{P_{2,\sigma_{1}\alpha_{1}\alpha_{2}}^{(2)}\left(\theta,\epsilon_{1},\epsilon_{2}\right)}{\epsilon_{2}}+I_{\sigma_{2}-\alpha_{1}-\alpha_{2}}^{(2)}\left(\theta,\epsilon_{1},\epsilon_{2}\right)\Big)\label{eq:Fssaaeps}
\end{eqnarray}
Each of the two integrals can now be written as a real integral and
is finite whenever $\epsilon_{1,2}\to0$. On the other hand, they
can have $O(\epsilon_{1,2})$ contributions. We separate each term
in the sum and the various orders in $\epsilon_{1,2}$ by writing
\begin{eqnarray*}
I_{\sigma\alpha_{1}\alpha_{2}}^{(j)}\left(\theta,\epsilon_{1},\epsilon_{2}\right) & = & I_{\sigma\alpha_{1}\alpha_{2}}^{(j)}\left(\theta\right)+\epsilon_{1}J_{\sigma\alpha_{1}\alpha_{2}}^{(j,1)}\left(\theta\right)+\epsilon_{2}J_{\sigma\alpha_{1}\alpha_{2}}^{(j,2)}\left(\theta\right)
\end{eqnarray*}
where $\alpha_{1,2}=\pm$ label the terms in the sum (\ref{eq:deompIss}),
while $\sigma=\pm$ labels the contour index in the sum (\ref{eq:FZboson}).
The prefactor $\mathcal{A}$ can be subjected to the same analysis.
Its finite part and $O(\epsilon_{1,2})$ contributions can be isolated
as

\begin{eqnarray}
\mathcal{A}_{ssaa}(\theta_{1}+i\pi+\epsilon_{1},\theta_{2}+i\pi+\epsilon_{2},\theta_{2},\theta_{1}) & = & \mathcal{A}_{ssaa}+\epsilon_{1}\mathcal{A}_{ssaa,1}+\epsilon_{2}\mathcal{A}_{ssaa,2}+\epsilon_{1}\epsilon_{2}\mathcal{A}_{ssaa,12}\label{eq:AssaaX}
\end{eqnarray}

The same reasoning can be applied to the form factors in which particles
of opposite charge are adjacent. In this case we have:
\begin{eqnarray}
 &  & F_{sasa}\left(\vartheta_{1}+i\pi+\epsilon_{1},\vartheta_{2}+i\pi,\vartheta_{2}+\epsilon_{2},\vartheta_{1}\right)=\nonumber \\
 &  & -\mathcal{G}_{k}\mathcal{A}_{sasa}\left(\theta,\epsilon_{1},\epsilon_{2}\right)\sum_{\sigma_{1}\sigma_{2}}\sigma_{1}e^{i\sigma_{1}\pi(1+1/\xi)/2}\sigma_{2}e^{i\sigma_{2}\pi(1+1/\xi)/2}\sum_{\alpha_{1}\alpha_{2}}\alpha_{1}\alpha_{2}\nonumber \\
 &  & \left(\frac{P_{1;\sigma_{1}\alpha_{1}\alpha_{2}}^{(1)}\left(\theta,\epsilon_{1},\epsilon_{2}\right)}{\epsilon_{1}}+\frac{P_{2;\sigma_{1}\alpha_{1}\alpha_{2}}^{(1)}\left(\theta,\epsilon_{1},\epsilon_{2}\right)}{\epsilon_{2}}+I_{\sigma_{1}\alpha_{1}\alpha_{2}}^{(1)}\left(\theta,\epsilon_{1},\epsilon_{2}\right)\right)\nonumber \\
 &  & \left(\frac{P_{1;\sigma_{2}\alpha_{1}\alpha_{2}}^{(2)}\left(\theta,\epsilon_{1},\epsilon_{2}\right)}{\epsilon_{1}}+\frac{P_{2;\sigma_{2}\alpha_{1}\alpha_{2}}^{(2)}\left(\theta,\epsilon_{1},\epsilon_{2}\right)}{\epsilon_{2}}+I_{\sigma_{2}-\alpha_{1}-\alpha_{2}}^{(2)}\left(\theta,\epsilon_{1},\epsilon_{2}\right)\right)\label{eq:Fsasaeps}
\end{eqnarray}
where we introduced the parametrization

\begin{eqnarray*}
I_{\sigma_{1}\alpha_{1}\alpha_{2}}^{(j)}\left(\theta,\epsilon_{1},\epsilon_{2}\right) & = & I_{\sigma\alpha_{1}\alpha_{2}}^{(j)}\left(\theta\right)+\epsilon_{1}K_{\sigma_{1}\alpha_{1}\alpha_{2}}^{(j,1)}\left(\theta\right)+\epsilon_{2}K_{\sigma_{1}\alpha_{1}\alpha_{2}}^{(j,2)}\left(\theta\right)
\end{eqnarray*}
and
\begin{eqnarray}
\mathcal{A}_{sasa}(\theta_{1}+i\pi+\epsilon_{1},\theta_{2}+i\pi,\theta_{2}+\epsilon_{2},\theta_{1}) & = & \mathcal{A}_{sasa}+\epsilon_{1}\mathcal{A}_{sasa,1}+\epsilon_{2}\mathcal{A}_{sasa,2}+\epsilon_{1}\epsilon_{2}\mathcal{A}_{sasa,12}\label{eq:AsasaX}
\end{eqnarray}
Note that in this expression, due to the presence of an integration
in the (\ref{eq:FZboson}), it is more convenient to expand around
the antisoliton rapidity.

The only singularities are coming from the residues picked up from
the principal poles of the $W$ functions, during the process of deforming
the contours, while the remaining integrals are regular. Note that
because of the presence of two antisolitons, each of the contours
generates a singularity. However, the contraction of the vertexes
$\left\langle \left\langle e^{-i\bar{\phi}(\vartheta_{1})}e^{-i\bar{\phi}(\vartheta_{2})}\right\rangle \right\rangle $
is zero for coinciding rapidities, hence the poles at $\epsilon_{1}\to0$
and $\epsilon_{2}\to0$ are simple.

\subsection{The integral parts}

To regularize the integral representation, we borrow a method from
the original paper \cite{Lukyanov:1993pn} and note that the integrals
associated to antisolitons can be interpreted as the analytic continuation
of Fourier transforms to imaginary arguments. Since the diagonal part
of the form factor written in terms of integrals over hyperbolic functions
only, one is able to compute the Fourier transforms 
\begin{equation}
\hat{f}(z)=\intop\frac{d\gamma}{2\pi}e^{iz\gamma}f(\gamma)
\end{equation}
explicitly, and then the resulting convolution can be evaluated numerically.

Let us define the functions:

\begin{eqnarray}
C(\gamma,\theta)^{-1} & = & \cosh(\gamma+\theta/2)\cosh(\gamma-\theta/2)\nonumber \\
S_{\alpha\beta}(\gamma,\theta)^{-1} & = & \text{\ensuremath{\sinh}}\frac{\gamma+\theta/2+i\alpha\pi/2}{\xi}\text{\ensuremath{\sinh}}\frac{\gamma-\theta/2+i\beta\pi/2}{\xi}
\end{eqnarray}
which we can use to write
\begin{equation}
\frac{1}{\bar{G}(\gamma+\frac{\theta}{2}-i\pi+i\alpha\frac{\pi}{2})\bar{G}(\gamma-\frac{\theta}{2}-i\pi+i\beta\frac{\pi}{2})}=\left(\frac{4}{\xi\mathcal{C}_{2}}\right)^{2}C(\gamma,\theta)S_{\alpha\beta}(\gamma,\theta)
\end{equation}
Their Fourier transforms are
\begin{equation}
\hat{C}(z,\theta)=\frac{\sin\frac{\theta z}{2}}{\sinh\frac{\pi z}{2}\sinh\theta}
\end{equation}
\begin{equation}
\hat{S}_{++}(z,\theta)=\frac{-\xi\sin\frac{z\theta}{2}e^{-\frac{\pi(\xi-1)z}{2}-\pi z\xi\left\lfloor \frac{1}{2\xi}\right\rfloor }}{\sinh\frac{\pi z\xi}{2}\sinh\frac{\theta}{\xi}}\qquad\hat{S}_{+-}(z,\theta)=\frac{i\xi\sinh\left(\frac{(\xi-1)\pi-i\theta}{2}+\pi\xi\left\lfloor \frac{1}{2\xi}\right\rfloor \right)z}{\sinh\frac{\pi z\xi}{2}\sinh\frac{\theta+i\pi}{\xi}}
\end{equation}
with the obvious symmetries
\begin{equation}
\hat{S}_{\alpha,\beta}(-z,\theta)=\hat{S}_{-\beta,-\alpha}(z,\theta)\qquad\hat{S}_{\alpha,\beta}(z,-\theta)=\hat{S}_{\beta,\alpha}(z,\theta)
\end{equation}
On the other hand, the Fourier transform of the logarithmic derivative
of the $W$ function is easily obtained as

\begin{equation}
\hat{L}(z,\theta)=\intop\frac{d\gamma}{2\pi}e^{iz\gamma}\partial_{\gamma}\log W(\gamma-i\pi)=\frac{1}{2i}\left(\frac{1}{\sinh\frac{\pi z}{2}}-\frac{\sinh\frac{\pi(\xi-1)z}{2}}{\sinh\pi z\sinh\frac{\xi\pi z}{2}}\right)e^{iz\theta/2}
\end{equation}
and also, using the definitions (\ref{eq:defGbar})
\begin{equation}
\hat{L}_{\pm}(z,\theta)=\intop\frac{d\gamma}{2\pi}e^{iz\gamma}\partial_{\gamma}\log W(\gamma-\frac{\theta}{2}-i(1\mp1)\pi)=-\hat{L}(z)-\frac{1}{2i}\left(\frac{1}{\sinh\frac{\pi z}{2}}+\frac{e^{\mp\frac{\pi(\xi-1)z}{2}}}{\sinh\frac{\xi\pi z}{2}}\right)e^{i\frac{z\theta}{2}}
\end{equation}
Using this procedure, one obtains the integral part of the diagonal
$\mathcal{F}_{sa}$ form factors in the following form:
\begin{eqnarray}
I_{\sigma\alpha_{1}\alpha_{2}}^{(1)}(\theta)=I_{\sigma\alpha_{1}\alpha_{2}}^{(2)}(-\theta) & = & \sigma\left(\frac{4}{\xi\mathcal{C}_{2}}\right)^{2}\int_{\mathbb{R}}dx\hat{C}\left(i\left(2a+\alpha_{1}+\frac{1+\alpha_{2}}{\xi}\right)-x\right)\hat{S}_{-\sigma,+}(x,\theta)\nonumber \\
J_{\sigma\alpha_{1}\alpha_{2}}^{(1,1)}(\theta)=J_{\sigma\alpha_{1}\alpha_{2}}^{(2,2)}(-\theta) & = & -\sigma\left(\frac{4}{\xi\mathcal{C}_{2}}\right)^{2}\int_{\mathbb{R}}dx\hat{C}\left(i\left(2a+\alpha_{1}+\frac{1+\alpha_{2}}{\xi}\right)-x\right)\nonumber \\
 &  & \times\int_{\mathbb{R}}dy\hat{S}_{-\sigma,+}(x-y,\theta)\hat{L}(y,-\theta)\nonumber \\
J_{\sigma\alpha_{1}\alpha_{2}}^{(1,2)}(\theta)=J_{\sigma\alpha_{1}\alpha_{2}}^{(2,1)}(-\theta) & = & -\sigma\left(\frac{4}{\xi\mathcal{C}_{2}}\right)^{2}\int_{\mathbb{R}}dx\hat{C}\left(i\left(2a+\alpha_{1}+\frac{1+\alpha_{2}}{\xi}\right)-x\right)\nonumber \\
 &  & \times\int_{\mathbb{R}}dy\hat{S}_{-\sigma,+}(x-y,\theta)\hat{L}_{+}(y,\theta)
\end{eqnarray}
For the $\mathcal{F}_{ss}$ , we have instead:

\begin{eqnarray}
I_{\sigma\alpha_{1}\alpha_{2}}^{(1)}(\theta)=I_{-\sigma\alpha_{1}\alpha_{2}}^{(2)}(-\theta)^{*} & = & \sigma\int_{\mathbb{R}}dx\hat{C}\left(i(2a+\alpha_{1}+\frac{1+\alpha_{2}}{\xi})-x\right)\hat{S}_{-\sigma,+}(x,\theta)\nonumber \\
K_{\sigma\alpha_{1}\alpha_{2}}^{(1,1)}(\theta)=K_{\sigma\alpha_{1}\alpha_{2}}^{(1,2)}(-\theta) & = & -\sigma\int_{\mathbb{R}}dx\hat{C}\left(i(2a+\alpha_{1}+\frac{1+\alpha_{2}}{\xi})-x\right)\nonumber \\
 &  & \times\int_{\mathbb{R}}dy\hat{S}_{-\sigma,+}(x-y,\theta)\hat{L}(y,\theta)\nonumber \\
K_{\sigma\alpha_{1}\alpha_{2}}^{(2,2)}(\theta)=K_{\sigma\alpha_{1}\alpha_{2}}^{(2,1)}(-\theta) & = & K_{-\sigma\alpha_{1}\alpha_{2}}^{(1,1)}(-\theta)^{*}
\end{eqnarray}

\subsection{Poles\label{sub:Poles}}

Here we provide the formulas for the poles contribution. We write
the contributions from the poles associated to the deformation of
the contour on which the variable $\gamma_{j}$ is integrated ($j=1,2$)
as $\frac{1}{\epsilon_{k}}P_{k}^{(j)}\left(\theta,\epsilon_{1},\epsilon_{2}\right)=\frac{1}{\epsilon_{k}}P_{k}^{(j)}+P_{k;k}^{(j)}+\frac{\epsilon_{3-k}}{\epsilon_{k}}P_{k;3-k}^{(j)}$
with the index $k=1,2$ labeling the singularity and $\theta=\vartheta_{2}-\vartheta_{1}$.
For practical reasons, it will be more convenient to include the all
$O\left(\epsilon^{0}\right)$ contributions in the integral part,
which is easily done by shifting $I_{\sigma\alpha_{1}\alpha_{2}}^{(j)}\to I_{\sigma\alpha_{1}\alpha_{2}}^{(j)}+P_{1,1;\sigma\alpha_{1}\alpha_{2}}^{(j)}+P_{2,2;\sigma\alpha_{1}\alpha_{2}}^{(j)}$.

For the $\mathcal{F}_{ss}=\mathcal{F}_{aa}$ matrix element we find:
\begin{eqnarray}
P_{1;\sigma,\alpha_{1},\alpha_{1}}^{(1)}\left(\theta,\epsilon_{1},\epsilon_{2}\right) & = & i\sigma\hat{W}\left(-i\frac{\pi}{2}\right)^{2}\frac{e^{(A+\alpha_{1}+\alpha_{2}/\xi)(-\theta+i\sigma\pi)/2}}{\bar{G}(\theta-i\sigma\frac{\pi}{2}-i\frac{3\pi}{2})}\Big(1+\epsilon_{2}\partial_{\theta}\log W\left(\theta-i\sigma\frac{\pi}{2}-2\pi i\right)\nonumber \\
 &  & -\sigma\epsilon_{1}\partial_{\theta}\log\hat{W}\left(-i\frac{\pi}{2}\right)\Big)\nonumber \\
P_{2;\sigma,\alpha_{1},\alpha_{1}}^{(1)}\left(\theta,\epsilon_{1},\epsilon_{2}\right) & = & i\hat{W}\left(-i\frac{\pi}{2}\right)^{2}\frac{e^{(A+\alpha_{1}+\alpha_{2}/\xi)(\theta-i\pi)/2}}{\bar{G}(\theta-i\sigma\frac{\pi}{2}-i\frac{3\pi}{2})}\left(1-\epsilon_{1}\partial_{\theta}\log W\left(\theta-i\frac{3\pi}{2}\right)\right)\nonumber \\
P_{2;\sigma,\alpha_{1},\alpha_{1}}^{(2)}\left(\theta,\epsilon_{1},\epsilon_{2}\right) & = & i\sigma\hat{W}\left(-i\frac{\pi}{2}\right)^{2}\frac{e^{(A+\alpha_{1}+\alpha_{2}/\xi)(\theta+i\sigma\pi)/2}}{\bar{G}(\theta+i\sigma\frac{\pi}{2}-i\frac{3\pi}{2})}\Big(1-\epsilon_{1}\partial_{\theta}\log W\left(\theta+i\sigma\frac{\pi}{2}\right)\nonumber \\
 &  & -\sigma\epsilon_{2}\partial_{\theta}\log\hat{W}\left(-i\frac{\pi}{2}\right)\Big)\nonumber \\
P_{1;\sigma,\alpha_{1},\alpha_{1}}^{(2)}\left(\theta,\epsilon_{1},\epsilon_{2}\right) & = & i\hat{W}\left(-i\frac{\pi}{2}\right)^{2}\frac{e^{(A+\alpha_{1}+\alpha_{2}/\xi)(-\theta-i\pi)/2}}{\bar{G}(\theta+i\sigma\frac{\pi}{2}-i\frac{3\pi}{2})}\left(1+\epsilon_{2}\partial_{\theta}\log W\left(\theta-i\frac{3\pi}{2}\right)\right)
\end{eqnarray}
again, having also included contour and expansion of the $\bar{G}$
function indexes. The notation $\partial_{\theta}\log\hat{W}\left(-i\pi/2\right)$
indicates the logarithmic derivative of the function $\hat{W}$ with
respect to the rapidity argument, evaluated at $\theta=-i\pi/2$.

For the $\mathcal{F}_{sa}$ form factor we obtain, instead
\begin{eqnarray}
P_{1;\sigma,\alpha_{1},\alpha_{1}}^{(1)}\left(\theta,\epsilon_{1},\epsilon_{2}\right) & = & i\sigma\hat{W}\left(-i\frac{\pi}{2}\right)^{2}\frac{e^{(A+\alpha_{1}+\alpha_{2}/\xi)(-\theta+i\sigma\pi)/2}}{\bar{G}(\theta-i\sigma\frac{\pi}{2}-i\frac{3\pi}{2})}\Big(1+\epsilon_{2}\partial_{\theta}\log W\left(\theta-i\sigma\frac{\pi}{2}-\pi i\right)\nonumber \\
 &  & -\sigma\epsilon_{1}\partial_{\theta}\log\hat{W}\left(-i\frac{\pi}{2}\right)\Big)\nonumber \\
P_{2;\sigma,\alpha_{1},\alpha_{1}}^{(1)}\left(\theta,\epsilon_{1},\epsilon_{2}\right) & = & i\hat{W}\left(-i\frac{\pi}{2}\right)^{2}\frac{e^{(A+\alpha_{1}+\alpha_{2}/\xi)(\theta-i\pi+\epsilon_{2})/2}}{\bar{G}(\theta-i\sigma\frac{\pi}{2}-i\frac{3\pi}{2}+\epsilon_{2})}\nonumber \\
 &  & \times\left(1-\epsilon_{1}\partial_{\theta}\log W\left(\theta-i\frac{3\pi}{2}+\epsilon_{2}\right)\right)\nonumber \\
P_{2;\sigma,\alpha_{1},\alpha_{1}}^{(2)}\left(\theta,\epsilon_{1},\epsilon_{2}\right) & = & i\sigma\hat{W}\left(-i\frac{\pi}{2}\right)^{2}\frac{e^{(A+\alpha_{1}+\alpha_{2}/\xi)(\theta+i\sigma\pi)/2}}{\bar{G}(\theta+i\sigma\frac{\pi}{2}-i\frac{3\pi}{2})}\Big(1-\epsilon_{1}\partial_{\theta}\log W\left(\theta+i\sigma\frac{\pi}{2}-i\pi\right)\nonumber \\
 &  & -\sigma\epsilon_{2}\partial_{\theta}\log\hat{W}\left(-i\frac{\pi}{2}\right)\Big)\nonumber \\
P_{1;\sigma,\alpha_{1},\alpha_{1}}^{(2)}\left(\theta,\epsilon_{1},\epsilon_{2}\right) & = & i\hat{W}\left(-i\frac{\pi}{2}\right)^{2}\frac{e^{(A+\alpha_{1}+\alpha_{2}/\xi)(-\theta-i\pi+\epsilon_{1})/2}}{\bar{G}(\theta+i\sigma\frac{\pi}{2}-i\frac{\pi}{2}-\epsilon_{1})}\nonumber \\
 &  & \times\left(1+\epsilon_{2}\partial_{\theta}\log W\left(\theta-i\frac{3\pi}{2}-\epsilon_{1}\right)\right)
\end{eqnarray}
from which the various orders can be easily extracted.

\subsection{Collecting the finite contributions}

Finally, we write the factor $\mathcal{A}_{sasa}$ from the contraction
(\ref{eq:Asasa0}) as in (\ref{eq:AsasaX}): 
\begin{eqnarray}
\mathcal{A}_{sasa} & = & \bar{G}(\theta_{21}-i\pi)\nonumber \\
 &  & \times\Bigg\{1-\epsilon_{1}\partial_{\theta_{21}}\log W(\theta_{21}-i\frac{3\pi}{2})-\epsilon_{2}\partial_{\theta_{21}}\log W(\theta_{21}-i\frac{\pi}{2})\nonumber \\
 &  & -\epsilon_{1}\epsilon_{2}\left(\partial_{\theta_{21}}\log W(\theta_{21}-i\frac{\pi}{2})\partial_{\theta_{21}}\log W(\theta_{21}-i\frac{3\pi}{2})+\partial_{\theta_{21}}^{2}\log G(\theta_{21}-i\pi)\right)\Bigg\}\label{eq:Asasa}
\end{eqnarray}
Substituting into (\ref{eq:Fsasaeps}) and selecting the order $O(\epsilon_{1,2}^{0})$
results in 

\begin{eqnarray}
\mathcal{F}_{sa}^{k,(c)}\left(\theta\right) & = & -\mathcal{G}_{k}\sum_{\sigma_{1}\sigma_{2}}\sigma_{1}e^{i\sigma_{1}\pi(1+1/\xi)/2}\sigma_{2}e^{i\sigma_{2}\pi(1+1/\xi)/2}\sum_{\alpha_{1}\alpha_{2}}\alpha_{1}\alpha_{2}\Big[\mathcal{A}\Big(P_{1;\sigma_{1}\alpha_{1}\alpha_{2}}^{(1)}J_{\sigma_{1}\alpha_{1}\alpha_{2}}^{(2,1)}\nonumber \\
 &  & +P_{2;\sigma_{1}\alpha_{1}\alpha_{2}}^{(1)}J_{\sigma_{1}\alpha_{1}\alpha_{2}}^{(2,2)}+J_{\sigma_{1}\alpha_{1}\alpha_{2}}^{(1,1)}P_{1;\sigma_{2}\alpha_{1}\alpha_{2}}^{(2)}+J_{\sigma_{1}\alpha_{1}\alpha_{2}}^{(1,2)}P_{2;\sigma_{2}\alpha_{1}\alpha_{2}}^{(2)}+I_{\sigma_{1}\alpha_{1}\alpha_{2}}^{(1)}I_{\sigma\alpha_{1}\alpha_{2}}^{(2)}\nonumber \\
 &  & +P_{1,2;\sigma_{1}\alpha_{1}\alpha_{2}}^{(1)}P_{2,1;\sigma_{2}\alpha_{1}\alpha_{2}}^{(2)}+P_{2,1;\sigma_{1}\alpha_{1}\alpha_{2}}^{(1)}P_{1,2;\sigma_{2}\alpha_{1}\alpha_{2}}^{(2)}\Big)+\mathcal{A}_{1}\left(P_{1;\sigma_{1}\alpha_{1}\alpha_{2}}^{(1)}I_{\sigma\alpha_{1}\alpha_{2}}^{(2)}+I_{\sigma_{1}\alpha_{1}\alpha_{2}}^{(1)}P_{1;\sigma_{2}\alpha_{1}\alpha_{2}}^{(2)}\right)\nonumber \\
 &  & +\mathcal{A}_{2}\left(P_{2;\sigma_{1}\alpha_{1}\alpha_{2}}^{(1)}I_{\sigma\alpha_{1}\alpha_{2}}^{(2)}+I_{\sigma_{1}\alpha_{1}\alpha_{2}}^{(1)}P_{2;\sigma_{2}\alpha_{1}\alpha_{2}}^{(2)}\right)+\mathcal{A}_{12}\left(P_{1;\sigma_{1}\alpha_{1}\alpha_{2}}^{(1)}P_{2;\sigma_{2}\alpha_{1}\alpha_{2}}^{(2)}+P_{2;\sigma_{1}\alpha_{1}\alpha_{2}}^{(1)}P_{1;\sigma_{2}\alpha_{1}\alpha_{2}}^{(2)}\right)\Big]\nonumber \\
\label{eq:Fsasaconn}
\end{eqnarray}
with

\begin{eqnarray}
\mathcal{A}=\mathcal{A}_{sasa} & = & \bar{G}(\theta-i\pi)\nonumber \\
\mathcal{A}_{1}=\mathcal{A}_{sasa,1}=\mathcal{A}_{2}^{*}=\mathcal{A}_{sasa,2} & = & -\partial_{\theta}\log W(\theta-i\frac{3\pi}{2})\nonumber \\
\mathcal{A}_{12}=\mathcal{A}_{sasa,12} & = & -\left|\partial_{\theta}\log W(\theta-i\frac{3\pi}{2})\right|^{2}-\partial_{\theta}^{2}\log G(\theta-i\pi)
\end{eqnarray}
and the functions $P_{k}^{(j)}$ from section \ref{sub:Poles}.

Using this result we have checked explicitly that $\mathcal{F}_{as}$
can be obtained by complex conjugation, as it is, possible to put
each term in (\ref{eq:Fsasaconn}) in correspondence with the terms
in $\mathcal{F}_{as}$.

For the case of $\mathcal{F}_{ss}$ , writing the contraction $\mathcal{A}_{ssaa}$
given in (\ref{eq:Assaa0}) in the form (\ref{eq:AssaaX}) gives:
\begin{eqnarray}
\mathcal{A}_{ssaa}(\theta_{1}+i\pi+\epsilon_{1},\theta_{2}+i\pi+\epsilon_{2},\theta_{2},\theta_{1}) & = & \bar{G}(\theta_{21}-i\pi)\nonumber \\
 &  & \times\Bigg\{1+\epsilon_{1}\partial_{\theta_{21}}\log W(\theta_{21}-i\frac{3\pi}{2})-\epsilon_{2}\partial_{\theta_{21}}\log W(\theta_{21}-i\frac{3\pi}{2})\nonumber \\
 &  & -\epsilon_{1}\epsilon_{2}\left(\partial_{\theta_{21}}\log W(\theta_{21}-i\frac{3\pi}{2})^{2}+\partial_{\theta_{21}}^{2}\log G(\theta_{21})\right)\Bigg\}\label{eq:Assaa}
\end{eqnarray}
Multiplying this contribution with the integral part as in (\ref{eq:Fssaaeps})
and selecting the order $\epsilon_{1,2}^{0}$, one obtains:
\begin{eqnarray}
\mathcal{F}_{ss}^{k(c)}\left(\theta\right) & = & \mathcal{G}_{k}\sum_{\sigma_{1}\sigma_{2}}\sigma_{1}e^{i\sigma_{1}\pi(1+1/\xi)/2}\sigma_{2}e^{i\sigma_{2}\pi(1+1/\xi)/2}\sum_{\alpha_{1}\alpha_{2}}\alpha_{1}\alpha_{2}e^{i\alpha_{2}\pi/\xi}\Big[\mathcal{A}\Big(P_{1;\sigma_{1}\alpha_{1}\alpha_{2}}^{(1)}K_{\sigma_{1}\alpha_{1}\alpha_{2}}^{(2,1)}\nonumber \\
 &  & +P_{2;\sigma_{2}\alpha_{1}\alpha_{2}}^{(1)}K_{\sigma_{1}\alpha_{1}\alpha_{2}}^{(2,2)}+K_{\sigma_{1}\alpha_{1}\alpha_{2}}^{(1,1)}P_{1;\sigma_{1}\alpha_{1}\alpha_{2}}^{(1)}+K_{\sigma_{1}\alpha_{1}\alpha_{2}}^{(1,2)}P_{2;\sigma_{2}\alpha_{1}\alpha_{2}}^{(2)}+I_{\sigma_{1}\alpha_{1}\alpha_{2}}^{(1)}I_{\sigma\alpha_{1}\alpha_{2}}^{(2)}\nonumber \\
 &  & +P_{1,2;\sigma_{1}\alpha_{1}\alpha_{2}}^{(1)}P_{2,1;\sigma_{2}\alpha_{1}\alpha_{2}}^{(2)}+P_{2,1;\sigma_{1}\alpha_{1}\alpha_{2}}^{(1)}P_{1,2;\sigma_{2}\alpha_{1}\alpha_{2}}^{(2)}\Big)+\mathcal{A}_{1}\Big(P_{1;\sigma_{1}\alpha_{1}\alpha_{2}}^{(1)}I_{\sigma\alpha_{1}\alpha_{2}}^{(2)}\nonumber \\
 &  & +I_{\sigma_{1}\alpha_{1}\alpha_{2}}^{(1)}P_{1;\sigma_{2}\alpha_{1}\alpha_{2}}^{(2)}\Big)+\mathcal{A}_{2}\left(P_{2;\sigma_{1}\alpha_{1}\alpha_{2}}^{(1)}I_{\sigma\alpha_{1}\alpha_{2}}^{(2)}+I_{\sigma_{1}\alpha_{1}\alpha_{2}}^{(1)}P_{2;\sigma_{2}\alpha_{1}\alpha_{2}}^{(2)}\right)\label{eq:Fssaaconn}\\
 &  & +\mathcal{A}_{12}\left(P_{1;\sigma_{1}\alpha_{1}\alpha_{2}}^{(1)}P_{2;\sigma_{2}\alpha_{1}\alpha_{2}}^{(2)}+P_{2;\sigma_{1}\alpha_{1}\alpha_{2}}^{(1)}P_{1;\sigma_{2}\alpha_{1}\alpha_{2}}^{(2)}\right)\Big]\nonumber 
\end{eqnarray}
where all the functions above depend on the rapidity difference $\theta=\theta_{1}-\theta_{2}$,
the pole functions are computed from section \ref{sub:Poles} and
\begin{eqnarray}
\mathcal{A}=\mathcal{A}_{ssaa} & = & \bar{G}(\theta-i\pi)\nonumber \\
\mathcal{A}_{1}=\mathcal{A}_{ssaa,1}=-\mathcal{A}_{2}=-\mathcal{A}_{ssaa,2} & = & \partial_{\theta}\log W(\theta-i\frac{3\pi}{2})\nonumber \\
\mathcal{A}_{12}=\mathcal{A}_{ssaa,12} & = & \partial_{\theta}\log W(\theta-i\frac{3\pi}{2})^{2}+\partial_{\theta}^{2}\log G(\theta)
\end{eqnarray}
The form factor $\mathcal{F}_{aa}$ can be formally obtained by the
substitution $\theta\rightarrow-\theta$, the latter being actually
a symmetry of the expression above, so $\mathcal{F}_{ss}=\mathcal{F}_{aa}$
as expected from charge conjugation symmetry. 

\bibliographystyle{utphys}
\bibliography{nondiagpaper}

\end{document}